%% file: main.tex
\pdfoutput=1
\documentclass[12pt,a4paper]{article}
\usepackage{ifthen} % for conditional statements
\newboolean{pdflatex}
\setboolean{pdflatex}{true} % False for eps figures 

\newboolean{articletitles}
\setboolean{articletitles}{true} % False removes titles in references

\newboolean{uprightparticles}
\setboolean{uprightparticles}{false} %True for upright particle symbols

\newboolean{inbibliography}
\setboolean{inbibliography}{false} %True once you enter the bibliography

\input{preamble}
\usepackage{longtable} % only for template; not usually to be used in PAPERs

\begin{document}

%%%%%%%%%%%%%%%%%%%%%%%%%
%%%%% Title     %%%%%%%%%
%%%%%%%%%%%%%%%%%%%%%%%%%
\renewcommand{\thefootnote}{\fnsymbol{footnote}}
\setcounter{footnote}{1}

% %%%%%%% CHOOSE TITLE PAGE--------
%\onecolumn
\input{title-LHCb-PAPER}

%\twocolumn
% %%%%%%%%%%%%% ---------

\renewcommand{\thefootnote}{\arabic{footnote}}
\setcounter{footnote}{0}

%%%%%%%%%%%%%%%%%%%%%%%%%%%%%%%%
%%%%%  Table of Content   %%%%%%
%%%%%%%%%%%%%%%%%%%%%%%%%%%%%%%%
%%%% Uncomment next 2 lines if desired
%\tableofcontents
%\cleardoublepage

%%%%%%%%%%%%%%%%%%%%%%%%%
%%%%% Main text %%%%%%%%%
%%%%%%%%%%%%%%%%%%%%%%%%%

\pagestyle{plain} % restore page numbers for the main text
\setcounter{page}{1}
\pagenumbering{arabic}

%% Uncomment during review phase. 
%% Comment before a final submission.
%\linenumbers

\input{introduction}

\input{observables}

\input{detector}

\input{selection}

\input{fit}

\input{results}

\input{interpretation}

\input{acknowledgements}

\addcontentsline{toc}{section}{References}
\setboolean{inbibliography}{true}
\bibliographystyle{LHCb}
\bibliography{main,LHCb-PAPER,LHCb-CONF,LHCb-DP,LHCb-TDR,bibliography}

\newpage
\input{LHCb_HD_authorlist_2015-02-27}

%\newpage
%\input{supplementary}

\end{document}

%% file: preamble.tex
\usepackage{microtype}
\usepackage{lineno}  % for line numbering during review
\usepackage{xspace} % To avoid problems with missing or double spaces after
\usepackage{caption} %these three command get the figure and table captions automatically small
\usepackage{enumerate}
\usepackage{placeins}

\usepackage{graphicx}  % to include figures (can also use other packages)
\usepackage{color}
\usepackage{colortbl}
\graphicspath{{./figs/}} % Make Latex search fig subdir for figures

\usepackage{amsmath} % Adds a large collection of math symbols
\usepackage{amssymb}
\usepackage{amsfonts}
\usepackage{upgreek} % Adds in support for greek letters in roman typeset

\newcommand*\patchAmsMathEnvironmentForLineno[1]{%
\expandafter\let\csname old#1\expandafter\endcsname\csname #1\endcsname
\expandafter\let\csname oldend#1\expandafter\endcsname\csname
end#1\endcsname
 \renewenvironment{#1}%
   {\linenomath\csname old#1\endcsname}%
   {\csname oldend#1\endcsname\endlinenomath}%
}
\newcommand*\patchBothAmsMathEnvironmentsForLineno[1]{%
  \patchAmsMathEnvironmentForLineno{#1}%
  \patchAmsMathEnvironmentForLineno{#1*}%
}
\AtBeginDocument{%
\patchBothAmsMathEnvironmentsForLineno{equation}%
\patchBothAmsMathEnvironmentsForLineno{align}%
\patchBothAmsMathEnvironmentsForLineno{flalign}%
\patchBothAmsMathEnvironmentsForLineno{alignat}%
\patchBothAmsMathEnvironmentsForLineno{gather}%
\patchBothAmsMathEnvironmentsForLineno{multline}%
\patchBothAmsMathEnvironmentsForLineno{eqnarray}%
}

\usepackage{hyperref}    % Hyperlinks in references
\usepackage[all]{hypcap} % Internal hyperlinks to floats.

\input{lhcb-symbols-def} % Add in the predefined LHCb symbols
\input{NewCommands} %New commands for hhpi0 analysis

\usepackage{cite} % Allows for ranges in citations
\usepackage{mciteplus}

%% file: lhcb-symbols-def.tex
\def\lhcb {\mbox{LHCb}\xspace}

\def\babar  {\mbox{BaBar}\xspace}

\def\MagUp {\mbox{\em Mag\kern -0.05em Up}\xspace}

\ifthenelse{\boolean{uprightparticles}}%
{
 
 \def\Pgamma      {\ensuremath{\upgamma}\xspace}

 \def\Ppi         {\ensuremath{\uppi}\xspace}

 \def\PDelta      {\ensuremath{\Delta}\xspace}                 
 \def\PXi      {\ensuremath{\Xi}\xspace}                 
 \def\PLambda      {\ensuremath{\Lambda}\xspace}                 
 \def\PSigma      {\ensuremath{\Sigma}\xspace}                 
 \def\POmega      {\ensuremath{\Omega}\xspace}                 
 \def\PUpsilon      {\ensuremath{\Upsilon}\xspace}                 
 
 \def\PB      {\ensuremath{\mathrm{B}}\xspace}                 
                  
 \def\PD      {\ensuremath{\mathrm{D}}\xspace}

 \def\PK      {\ensuremath{\mathrm{K}}\xspace}

 \def\Pb      {\ensuremath{\mathrm{b}}\xspace}                 
 \def\Pc      {\ensuremath{\mathrm{c}}\xspace}

 \def\Pi      {\ensuremath{\mathrm{i}}\xspace}

 \def\Ps      {\ensuremath{\mathrm{s}}\xspace}

}
{
 
 \def\Pgamma      {\ensuremath{\gamma}\xspace}

 \def\Ppi         {\ensuremath{\pi}\xspace}

 \mathchardef\PDelta="7101
 \mathchardef\PXi="7104
 \mathchardef\PLambda="7103
 \mathchardef\PSigma="7106
 \mathchardef\POmega="710A
 \mathchardef\PUpsilon="7107
                  
 \def\PB      {\ensuremath{B}\xspace}                 
                  
 \def\PD      {\ensuremath{D}\xspace}

 \def\PK      {\ensuremath{K}\xspace}

 \def\Pb      {\ensuremath{b}\xspace}                 
 \def\Pc      {\ensuremath{c}\xspace}

 \def\Pi      {\ensuremath{i}\xspace}

 \def\Ps      {\ensuremath{s}\xspace}

}

\makeatletter
\ifcase \@ptsize \relax% 10pt
  \newcommand{\miniscule}{\@setfontsize\miniscule{4}{5}}% \tiny: 5/6
\or% 11pt
  \newcommand{\miniscule}{\@setfontsize\miniscule{5}{6}}% \tiny: 6/7
\or% 12pt
  \newcommand{\miniscule}{\@setfontsize\miniscule{5}{6}}% \tiny: 6/7
\fi
\makeatother

\DeclareRobustCommand{\optbar}[1]{\shortstack{{\miniscule (\rule[.5ex]{1.25em}{.18mm})}
  \\ [-.7ex] $#1$}}

\def\g      {{\ensuremath{\Pgamma}}\xspace}

\def\squark    {{\ensuremath{\Ps}}\xspace}

\def\cquark    {{\ensuremath{\Pc}}\xspace}

\def\bquark    {{\ensuremath{\Pb}}\xspace}

\def\pion   {{\ensuremath{\Ppi}}\xspace}
\def\piz    {{\ensuremath{\pion^0}}\xspace}

\def\pip    {{\ensuremath{\pion^+}}\xspace}
\def\pim    {{\ensuremath{\pion^-}}\xspace}
\def\pipm   {{\ensuremath{\pion^\pm}}\xspace}
\def\pimp   {{\ensuremath{\pion^\mp}}\xspace}

\def\kaon    {{\ensuremath{\PK}}\xspace}
  \def\Kbar    {{\kern 0.2em\overline{\kern -0.2em \PK}{}}\xspace}

\def\KorKbar    {\kern 0.18em\optbar{\kern -0.18em K}{}\xspace}

\def\Kp      {{\ensuremath{\kaon^+}}\xspace}
\def\Km      {{\ensuremath{\kaon^-}}\xspace}
\def\Kpm     {{\ensuremath{\kaon^\pm}}\xspace}
\def\Kmp     {{\ensuremath{\kaon^\mp}}\xspace}

\def\Kstar   {{\ensuremath{\kaon^*}}\xspace}

  \def\Dbar    {{\kern 0.2em\overline{\kern -0.2em \PD}{}}\xspace}
\def\D       {{\ensuremath{\PD}}\xspace}
\def\Db      {{\ensuremath{\Dbar}}\xspace}
\def\DorDbar    {\kern 0.18em\optbar{\kern -0.18em D}{}\xspace}
\def\Dz      {{\ensuremath{\D^0}}\xspace}
\def\Dzb     {{\ensuremath{\Dbar{}^0}}\xspace}

\def\Dstar   {{\ensuremath{\D^*}}\xspace}

\def\Dstarpm {{\ensuremath{\D^{*\pm}}}\xspace}

\def\B       {{\ensuremath{\PB}}\xspace}
\def\Bbar    {{\ensuremath{\kern 0.18em\overline{\kern -0.18em \PB}{}}}\xspace}

\def\BorBbar    {\kern 0.18em\optbar{\kern -0.18em B}{}\xspace}

\def\Bu      {{\ensuremath{\B^+}}\xspace}
\def\Bub     {{\ensuremath{\B^-}}\xspace}
\def\Bp      {{\ensuremath{\Bu}}\xspace}
\def\Bm      {{\ensuremath{\Bub}}\xspace}

\def\Bmp     {{\ensuremath{\B^\mp}}\xspace}

\def\Bs      {{\ensuremath{\B^0_\squark}}\xspace}

\def\BsorBsbar    {\kern 0.02em\optbar{\kern -0.02em \B_\squark}{}\xspace}

  \def\Y#1S{\ensuremath{\PUpsilon{(#1S)}}\xspace}% no space before {...}!

\def\Lbar        {{\ensuremath{\kern 0.1em\overline{\kern -0.1em\PLambda}}}\xspace}
\def\LorLbar    {\kern 0.18em\optbar{\kern -0.18em \PLambda}{}\xspace}

\newcommand{\decay}[2]{\ensuremath{#1\!\to #2}\xspace}         % {\Pa}{\Pb \Pc}

\def\to                 {\ensuremath{\rightarrow}\xspace}

\def\CP                {{\ensuremath{C\!P}}\xspace}

\def\AT#1     {\ensuremath{A_{\mathrm{T}}^{#1}}\xspace}           % 2

\def\C#1      {\ensuremath{\mathcal{C}_{#1}}\xspace}                       % 9
\def\Cp#1     {\ensuremath{\mathcal{C}_{#1}^{'}}\xspace}                    % 7
\def\Ceff#1   {\ensuremath{\mathcal{C}_{#1}^{\mathrm{(eff)}}}\xspace}        % 9  
\def\Cpeff#1  {\ensuremath{\mathcal{C}_{#1}^{'\mathrm{(eff)}}}\xspace}       % 7
\def\Ope#1    {\ensuremath{\mathcal{O}_{#1}}\xspace}                       % 2
\def\Opep#1   {\ensuremath{\mathcal{O}_{#1}^{'}}\xspace}                    % 7

\newcommand{\tev}{\ifthenelse{\boolean{inbibliography}}{\ensuremath{~T\kern -0.05em eV}\xspace}{\ensuremath{\mathrm{\,Te\kern -0.1em V}}}\xspace}
\newcommand{\gev}{\ensuremath{\mathrm{\,Ge\kern -0.1em V}}\xspace}
\newcommand{\mev}{\ensuremath{\mathrm{\,Me\kern -0.1em V}}\xspace}
\newcommand{\kev}{\ensuremath{\mathrm{\,ke\kern -0.1em V}}\xspace}
\newcommand{\ev}{\ensuremath{\mathrm{\,e\kern -0.1em V}}\xspace}
\newcommand{\gevc}{\ensuremath{{\mathrm{\,Ge\kern -0.1em V\!/}c}}\xspace}
\newcommand{\mevc}{\ensuremath{{\mathrm{\,Me\kern -0.1em V\!/}c}}\xspace}
\newcommand{\gevcc}{\ensuremath{{\mathrm{\,Ge\kern -0.1em V\!/}c^2}}\xspace}
\newcommand{\gevgevcccc}{\ensuremath{{\mathrm{\,Ge\kern -0.1em V^2\!/}c^4}}\xspace}
\newcommand{\mevcc}{\ensuremath{{\mathrm{\,Me\kern -0.1em V\!/}c^2}}\xspace}

\def\mum  {\ensuremath{{\,\upmu\rm m}}\xspace}

\def\invfb   {\ensuremath{\mbox{\,fb}^{-1}}\xspace}

\newcommand{\chisq}{\ensuremath{\chi^2}\xspace}

\newcommand{\chisqip}{\ensuremath{\chi^2_{\rm IP}}\xspace}

\def\gsim{{~\raise.15em\hbox{$>$}\kern-.85em
          \lower.35em\hbox{$\sim$}~}\xspace}
\def\lsim{{~\raise.15em\hbox{$<$}\kern-.85em
          \lower.35em\hbox{$\sim$}~}\xspace}

\def\ptot       {\mbox{$p$}\xspace}
\def\pt         {\mbox{$p_{\rm T}$}\xspace}

\def\evtgen     {\mbox{\textsc{EvtGen}}\xspace}

\def\geant      {\mbox{\textsc{Geant4}}\xspace}

\def\photos     {\mbox{\textsc{Photos}}\xspace}

\def\pythia     {\mbox{\textsc{Pythia}}\xspace}

\def\tell1  {TELL1\xspace}
\def\ukl1   {UKL1\xspace}

%% file: NewCommands.tex
\def\BToDK               {\decay{\Bmp}{\D\Kmp}}
\def\BToDPi		 {\decay{\Bmp}{\D\pimp}}
\def\BToDh		{\decay{\Bmp}{\D h^{\mp}}}

\def\BToDHWithDToKPiPi0	{\decay{\Bmp}{[K\pi\piz]_Dh^{\mp}}}
\def\BToDKWithDToKPiPi0	{\decay{\Bmp}{[K\pi\piz]_D\Kmp}}
\def\BToDPiWithDToKPiPi0	{\decay{\Bmp}{[K\pi\piz]_D\pimp}}
\def\supBToDKWithDToKPiPi0	{\decay{\Bmp}{[\pimp\Kpm\piz]_D\Kmp}}
\def\favBToDKWithDToKPiPi0	{\decay{\Bmp}{[\Kmp\pipm\piz]_D\Kmp}}
\def\supBToDPiWithDToKPiPi0	{\decay{\Bmp}{[\pimp\Kpm\piz]_D\pimp}}
\def\favBToDPiWithDToKPiPi0	{\decay{\Bmp}{[\Kmp\pipm\piz]_D\pimp}}
\def\supBToDHWithDToKPiPi0	{\decay{\Bmp}{[\pimp\Kpm\piz]_Dh^{\mp}}}
\def\favBToDHWithDToKPiPi0	{\decay{\Bmp}{[\Kmp\pipm\piz]_Dh^\mp}}

\def\BToDHWithDToKKPi0	{\decay{\Bmp}{[\Kp\Km\piz]_Dh^{\mp}}}
\def\BToDKWithDToKKPi0 	{\decay{\Bmp}{[\Kp\Km\piz]_D\Kmp}}
\def\BToDPiWithDToKKPi0	{\decay{\Bmp}{[\Kp\Km\piz]_D\pimp}}
\def\BToDHWithDToPiPiPi0	{\decay{\Bmp}{[\pip\pim\piz]_Dh^{\mp}}}
\def\BToDKWithDToPiPiPi0	{\decay{\Bmp}{[\pip\pim\piz]_D\Kmp}}
\def\BToDPiWithDToPiPiPi0		{\decay{\Bmp}{[\pip\pim\piz]_D\pimp}}
\def\BToDHWithDToHHPi0 {\decay{\Bmp}{[h^+ h^- \piz]_D h^\mp}}

%% file: title-LHCb-PAPER.tex
% $Id: title-LHCb-PAPER.tex 67452 2015-02-10 11:22:35Z roldeman $
% ===============================================================================
% Purpose: LHCb-PAPER journal paper title page template
% Author: 
% Created on: 2010-09-25
% ===============================================================================

%%%%%%%%%%%%%%%%%%%%%%%%%
%%%%%  TITLE PAGE  %%%%%%
%%%%%%%%%%%%%%%%%%%%%%%%%
\begin{titlepage}
\pagenumbering{roman}

% Header ---------------------------------------------------
\vspace*{-1.5cm}
\centerline{\large EUROPEAN ORGANIZATION FOR NUCLEAR RESEARCH (CERN)}
\vspace*{1.5cm}
\noindent
\begin{tabular*}{\linewidth}{lc@{\extracolsep{\fill}}r@{\extracolsep{0pt}}}
\ifthenelse{\boolean{pdflatex}}% Logo format choice
{\vspace*{-2.7cm}\mbox{\!\!\!\includegraphics[width=.14\textwidth]{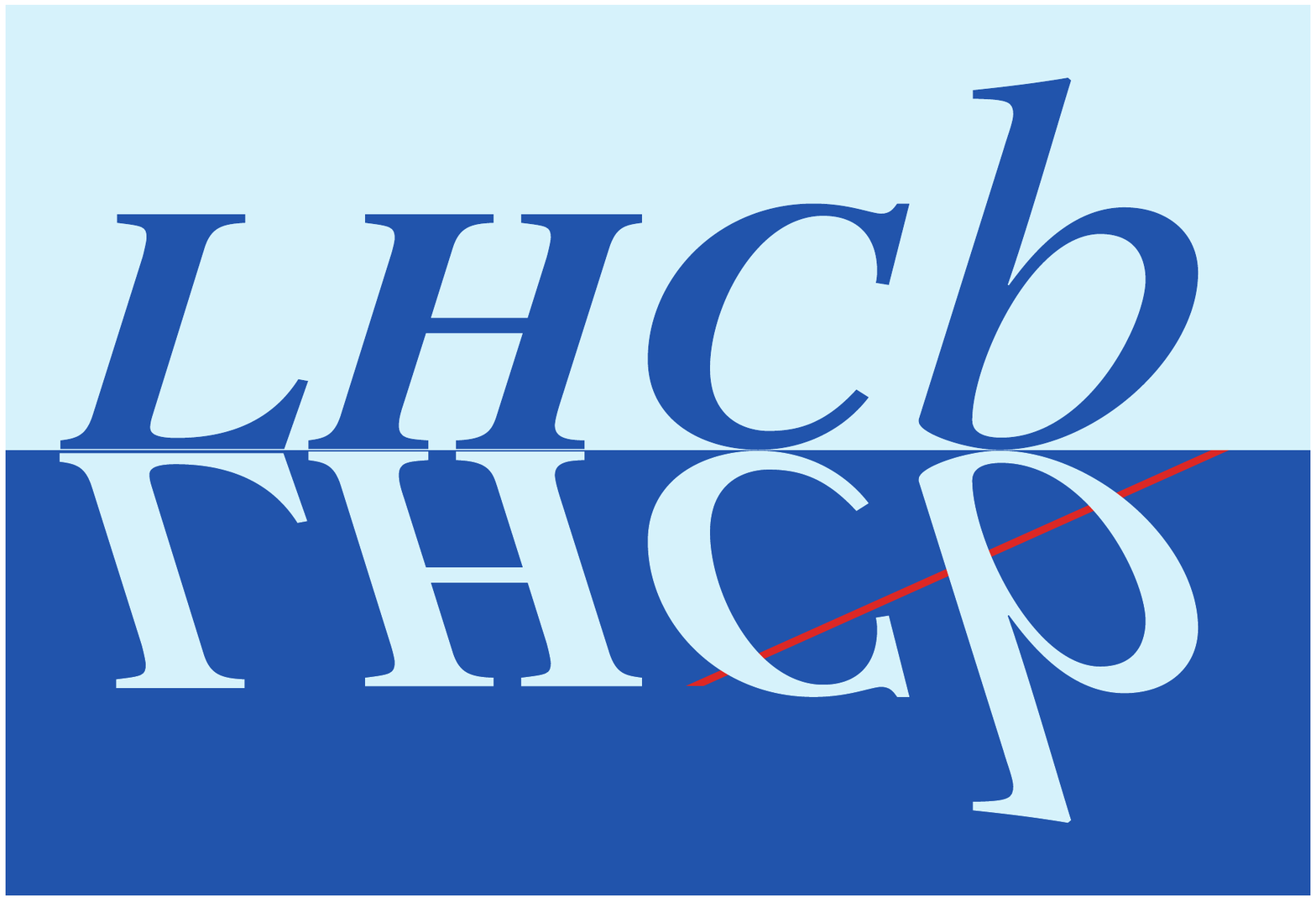}} & &}%
{\vspace*{-1.2cm}\mbox{\!\!\!\includegraphics[width=.12\textwidth]{lhcb-logo.eps}} & &}%
\\
 & & CERN-PH-EP-2015-097 \\  % ID 
 & & LHCb-PAPER-2015-014 \\  % ID 
 & & June 11, 2015 \\ % Date - Can also hardwire e.g.: 23 March 2010
\end{tabular*}

\vspace*{4.0cm}

% Title --------------------------------------------------
{\bf\boldmath\huge
\begin{center}
A study of $C\!P$ violation in $B^\mp \rightarrow Dh^\mp$ ($h=K,\pi$) with the modes $D \rightarrow K^\mp \pi^\pm \pi^0$, $D \rightarrow \pi^+\pi^-\pi^0$ and $D \rightarrow K^+K^-\pi^0$
\end{center}
}

\vspace*{1.0cm}

% Authors -------------------------------------------------
\begin{center}
%In the footnote, replace 'paper' by 'letter' in case of submission to PRL or PLB 
The LHCb collaboration\footnote{Authors are listed at the end of this paper.}
\end{center}

\vspace{\fill}

% Abstract -----------------------------------------------
\begin{abstract}
  \noindent
An analysis of the decays of $B^\mp \rightarrow D K^\mp$ and $B^\mp \rightarrow D \pi^\mp $ is presented in which the $D$ meson is reconstructed in the three-body final states $K^\mp \pi^\pm \pi^0$, $\pi^+ \pi^- \pi^0$ and $K^+ K^- \pi^0$.  Using data from LHCb corresponding to an integrated luminosity of 3.0\,\invfb of $pp$ collisions, measurements of several $C\!P$ observables are performed.  First observations are obtained of the suppressed ADS decay $B^\mp \rightarrow [\pi^\mp K^\pm \pi^0]_D \pi^\mp$ and the quasi-GLW decay $\B^\mp \rightarrow [K^+ K^- \pi^0]_D \pi^\mp$. The results are interpreted in the context of the unitarity triangle angle $\gamma$ and related parameters.
  
\end{abstract}

\vspace*{1.0cm}

\begin{center}
  Submitted to Phys.~Rev.~D
\end{center}

\vspace{\fill}

{\footnotesize 
\centerline{\copyright~CERN on behalf of the \lhcb collaboration, licence \href{http://creativecommons.org/licenses/by/4.0/}{CC-BY-4.0}.}}
\vspace*{2mm}

\end{titlepage}

%%%%%%%%%%%%%%%%%%%%%%%%%%%%%%%%
%%%%%  EOD OF TITLE PAGE  %%%%%%
%%%%%%%%%%%%%%%%%%%%%%%%%%%%%%%%

%  empty page follows the title page ----
\newpage
\setcounter{page}{2}
\mbox{~}

\cleardoublepage

%% file: introduction.tex
\section{Introduction}
\label{sec:Introduction}

Precise measurements of the parameters of the Cabibbo--Kobayashi--Maskawa unitarity triangle~\cite{Cabibbo:1963yz,*Kobayashi:1973fv}  are of great value in searching for manifestations of new physics in the flavour sector.  In particular, the determination of the angle $\gamma \equiv \arg({-V_{\rm ud}V^*_{\rm ub}/  V_{\rm cd}V_{\rm cb}^*})$   (also denoted as $\phi_3$ in the literature) in processes involving tree-level decays provides a Standard Model (SM) benchmark against which observables more sensitive to new physics contributions can be compared.   Currently such comparisons are limited by the uncertainty on $\gamma$, which is $\sim 7^\circ$~\cite{Lees:2013zd,Trabelsi:2013uj,LHCb-PAPER-2013-020,*LHCb-CONF-2014-004,PDG2014}. More precise measurements and new analysis strategies are therefore required.

Sensitivity to \g in tree-level processes may be obtained through the study of  \CP-violating observables in the decays $B^\mp \to \D h^\mp$,
where \D indicates a neutral charm meson which decays in a  mode common to both \Dz and \Dzb states, and $h$, the bachelor hadron, is either a kaon or a pion. In the case of $B^{-} \to \D K^{-}$, interference occurs between the suppressed $b \to u \bar{c} s$ and favoured $b \to c \bar{u} s$ quark-level transitions, and similarly for the charge-conjugate decay. The magnitude of the interference is governed by three parameters: the weak-phase difference, \g, the \CP-conserving strong-phase difference, $\delta_B$, and the ratio of the magnitudes of the two amplitudes, $r_B$. Similar interference effects occur in the case when the bachelor hadron is a pion, but then additional Cabibbo suppression factors mean that the sensitivity to \g is much reduced. Many classes of $D$ decay can be exploited. Important examples  include the so-called ADS modes~\cite{Atwood:1996ci,*Atwood:2000ck}, which are decays to quasi flavour eigenstates such as $D \to K^\mp\pi^\pm$, and the GLW modes~\cite{Gronau:1990ra,*Gronau:1991dp}, which are decays to \CP eigenstates such as $D \to K^+K^-$.  Measurements exist from LHCb that follow both the ADS and GLW approaches~\cite{LHCb-PAPER-2012-001,LHCb-PAPER-2012-055,LHCb-PAPER-2013-068,LHCb-PAPER-2014-028}, as well as alternative methods~\cite{LHCb-PAPER-2014-041,LHCb-PAPER-2014-017}.

In the case that the $D$ meson decays to three or more hadrons, the interference effects that are sensitive to $\gamma$ vary over the phase space of the $D$ decay due to the role of strongly-decaying intermediate resonances. If the $D$ decay is analysed inclusively, the integration over phase space in general dilutes the net sensitivity.  For multibody ADS modes the dilution factor can be measured with $D \Db$ pairs coherently produced at the $\psi(3770)$ resonance~\cite{COHERENCE}.  LHCb has previously made use of such measurements performed with data from the CLEO-c experiment~\cite{ Libby:2014rea,NORMLOWREY,Insler:2012pm} in  $B^\pm \to \D h^\pm$ analyses exploiting the modes $D \to K^\mp \pi^\pm \pi^-\pi^+$~\cite{LHCb-PAPER-2012-055} and  $D \to K^0_{\rm S} K^\mp \pi^\pm$~\cite{LHCb-PAPER-2013-068}.
It has recently been pointed out~\cite{Nayak:2014tea} that similar considerations apply to self-conjugate multibody modes such as $D \to \pi^+\pi^-\pi^0$.  These modes approximate to \CP eigenstates, and hence a \BToDK analysis that employs them can be considered a quasi-GLW (qGLW) analysis.  In this case the dilution factor is related to how closely the mode approaches a \CP eigenstate, and can also be measured at the open charm threshold.

This paper presents the measurement of \CP observables from $B^\pm \to \D h^\pm$ decays, where \D mesons are reconstructed using three different multibody final states.  These decays are the ADS channel $\D \to K^\mp \pi^\pm \piz$ and the quasi-GLW modes $\D \to \pip \pim \piz$ and $\D \to \Kp \Km \piz$.  In all cases, higher sensitivity is attained compared with the results of the previous measurements which exist from the \babar~\cite{Lees:2011up} and Belle~\cite{Nayak:2013tgg} collaborations for the ADS channel, and from \babar for the mode $\D \to \pip \pim \piz$~\cite{Aubert:2007ii}.  Measurements at the $\psi(3770)$ resonance~\cite{Libby:2014rea,NORMLOWREY,Nayak:2014tea} indicate that the dilution effects in $ D \to K^\mp \pi^\pm \pi^0$ and  $D \to \pi^+\pi^-\pi^0$ are rather small, making these decays particularly suitable for an inclusive analysis.

This paper is organised as follows.  Section~\ref{sec:observables} introduces the observables that the analysis seeks to measure, and explains how they are related to the underlying physics parameters and the dilution factors that are determined externally. Section~\ref{sec:Detector} describes the LHCb detector and the data set on which the analysis is based.  Sections~\ref{sec:Selection} and~\ref{sec:Fit} present the candidate selection and the analysis procedure. Results are given in Sect.~\ref{sec:Results}, together with a discussion of the systematic uncertainties.  In Sect.~\ref{sec:Interpretation} the measured observables are interpreted in terms of $\gamma$ and the other physics parameters, and conclusions are drawn.  

\

%% file: observables.tex
\section{Observables and external inputs}
\label{sec:observables}

In the ADS channel there exist two suppressed modes, $\Bmp \to [\pi^\mp K^\pm \pi^0]_D h^\mp$, and two favoured modes, $\Bmp \to [ K^\mp \pi^\pm \pi^0]_D h^\mp$, for $h= K$ and $\pi$.  In both cases the suppressed modes are as yet unobserved, although Belle has reported first evidence for $\Bmp \to [\pi^\mp K^\pm \pi^0]_D K^\mp$ and \supBToDPiWithDToKPiPi0 \cite{Nayak:2013tgg}.  As is customary in an ADS analysis, the ratio
\begin{equation}
R_{{\rm ADS}(h)}^{K \pi\pi^0} \equiv \frac{ \Gamma(\Bm \to [\pi^- \Kp \pi^0]_D h^-) + \Gamma(\Bp\to [\pi^+ \Km \pi^0]_D h^+) }
{ \Gamma(\Bm \to [\Km \pi^+  \pi^0]_D h^-) + \Gamma(\Bp\to [\Kp \pi^-  \pi^0]_D h^+) }
\label{eq:rads}
\end{equation}
is defined to give the relative rates of  the suppressed to the favoured decays.  The asymmetry
\begin{equation}
A_{{\rm ADS}(h)}^{K \pi\pi^0} \equiv \frac{ \Gamma(\Bm \to [\pi^- \Kp \pi^0]_D h^-) - \Gamma(\Bp\to [\pi^+ \Km \pi^0]_D h^+) }
{ \Gamma(\Bm \to [\pi^- \Kp \pi^0]_D h^-) + \Gamma(\Bp\to [\pi^+ \Km \pi^0]_D h^+) }
\label{eq:aads}
\end{equation}
quantifies the amount of \CP violation in the suppressed modes.  An asymmetry is also constructed for the favoured channels,
\begin{equation}
A_{K}^{K \pi\pi^0} \equiv \frac{ \Gamma(\Bm \to [\Km \pi^+  \pi^0]_D K^-) - \Gamma(\Bp\to [\Kp \pi^-  \pi^0]_D K^+) }
{ \Gamma(\Bm \to [\Km \pi^+  \pi^0]_D K^-) + \Gamma(\Bp\to [\Kp \pi^-  \pi^0]_D K^+) }.
\label{eq:afavoured}
\end{equation}

The observables $R_{{\rm ADS}(K)}^{K \pi\pi^0}$ and $A_{{\rm ADS}(K)}^{K \pi\pi^0}$ carry the highest sensitivity to the angle $\gamma$; they depend on the underlying physics parameters as
\begin{eqnarray}
R_{{\rm ADS}(K)}^{K\pi\pi^0} & \approx & (r_B)^2 + (r_D^{K\pi\pi^0})^2 + 2 \kappa_D^{K\pi\pi^0} r_B r_D^{K\pi\pi^0} \cos(\delta_B + \delta_D^{K\pi\pi^0})\cos \gamma,
\label{eq:rads_phys} \\
A_{{\rm ADS}(K)}^{K\pi\pi^0} & \approx & \left[ 2 \kappa_D^{K\pi\pi^0}r_B r_D^{K\pi\pi^0} \sin(\delta_B + \delta_D^{K\pi\pi^0})\sin\gamma \right ] / 
R_{{\rm ADS}(K)}^{K\pi\pi^0}.
\label{eq:aads_phys}
\end{eqnarray}
Here $r_D^{K\pi\pi^0} \sim 0.05$~\cite{PDG2014} is the ratio of the magnitudes of the doubly Cabibbo-suppressed and Cabibbo-favoured $D$ decay amplitudes and $\delta^{K\pi\pi^0}_D$ is the strong-phase difference between the amplitudes, averaged over phase space. 
The coherence factor $\kappa_D^{K\pi\pi^0}$ accounts for possible dilution effects in the interference arising from the contribution of the intermediate resonances in the $D$ decay~\cite{COHERENCE}.  Both $\delta^{K\pi\pi^0}_D$ and $\kappa_D^{K\pi\pi^0}$ have been measured with quantum-correlated $D \Db$ decays collected at the $\psi(3770)$ resonance by the CLEO-c experiment, and have been found to be $(164^{+20}_{-14})^\circ$ and $0.82 \pm 0.07$, respectively~\cite{Libby:2014rea}, where the phase-difference $\delta^{K\pi\pi^0}_D$ is given in the convention where $\CP |D^0 \rangle = | \Dzb \rangle$.  The relatively large value of $\kappa_D^{K\pi\pi^0}$ means that the dilution effects are small, and hence this decay is a promising mode to exploit for the measurement of $\gamma$.
Note that for reasons of clarity Eqs.~\ref{eq:rads_phys} and~\ref{eq:aads_phys} are restricted to terms of $\mathcal{O}\left((r_B)^2,(r_D^{K\pi\pi^0})^2,(r_B r_D^{K\pi\pi^0})\right)$, and the small effects of $D^0\Dzb$ mixing are omitted.  Full expressions may be found in Ref.~\cite{Rama:2013voa}.

In the quasi-GLW analysis of the two self-conjugate modes $D \to h^{\prime +} h^{\prime -} \pi^0$  ($h^\prime = K,\pi)$, observables are defined analogously to those used in the \CP-eigenstate case. The first of these is the  ratio of partial widths 
\begin{equation}
R_{{\rm qGLW}}^{h^\prime h^\prime \pi^0} \equiv \frac{\Gamma(\Bm \to D_{F_+^{h^\prime h^\prime \pi^0}} K^-)  +   \Gamma(\Bp \to D_{F_+^{h^\prime h^\prime \pi^0}} K^+)}{ \Gamma(\Bm \to D^0 K^-)  +   \Gamma(\Bp \to \Dzb K^+) },
\label{eq:rqglw}
\end{equation}
where $D_{F_+^{h^\prime h^\prime \pi^0}}$ signifies a \D meson with fractional \CP-even content $F_+^{h^\prime h^\prime \pi^0}$.  Both the numerator and the denominator of Eq.~\ref{eq:rqglw} involve \B meson partial widths only and have no dependence on the \D meson branching fractions.  In practice, therefore, $R_{{\rm qGLW}}^{h^\prime h^\prime \pi^0}$ is determined by forming the ratio of two more ratios,
\begin{eqnarray}
R_{{\rm qGLW}}^{h^\prime h^\prime \pi^0}  & \approx & R_{K/\pi}^{h^\prime h^\prime \pi^0} / R_{K/\pi}^{K \pi \pi^0}, \label{eq:rqglwapprox} \\
R_{K/\pi}^{h^\prime h^\prime \pi^0} & \equiv &  \frac{ \Gamma(\Bm \to [h^\prime h^\prime \pi^0]_D K^-)  +  \Gamma(\Bp \to [h^\prime h^\prime \pi^0]_D K^+)  }{ \Gamma(\Bm \to [h^\prime h^\prime \pi^0]_D \pi^-)  +  \Gamma(\Bp \to [h^\prime h^\prime \pi^0]_D \pi^+)  }, \label{eq:rkpi1} \\
R_{K/\pi}^{K \pi \pi^0} & \equiv &  \frac{ \Gamma(\Bm \to [K^- \pi^+ \pi^0]_D K^-)  +  \Gamma(\Bp \to [K^+\pi^- \pi^0]_D K^+)  }
{ \Gamma(\Bm \to [K^- \pi^+ \pi^0]_D \pi^-)  +  \Gamma(\Bp \to [K^+\pi^- \pi^0]_D \pi^+)  }, \label{eq:rkpi2}
\end{eqnarray}
where the approximate equality in Eq.~\ref{eq:rqglwapprox} acknowledges that very small interference effects in the $B^\mp \to D \pi^\mp$ decays specified in Eqs.~\ref{eq:rkpi1} and~\ref{eq:rkpi2} can be neglected.  This is a good assumption because the ratio between interfering amplitudes in \BToDPi decays is known to be very small~\cite{LHCb-PAPER-2013-020,*LHCb-CONF-2014-004}.  Furthermore, the ratio $R_{K/\pi}^{h^\prime h^\prime \pi^0} / R_{K/\pi}^{K \pi \pi^0}$ may be interpreted in terms of the underlying physics parameters, taking these interference effects into account.  Asymmetries, $A_{{\rm qGLW}(h)}^{h^\prime h^\prime \pi^0}$ ($h=K,\pi$), are also constructed, where
\begin{equation}
A_{{\rm qGLW}(h)}^{h^\prime h^\prime \pi^0} \equiv \frac{ \Gamma(\Bm \to [h^\prime h^\prime \pi^0]_D h^-) -  \Gamma(\Bp \to [h^\prime h^\prime \pi^0]_D h^+)  }
{ \Gamma(\Bm \to [h^\prime h^\prime \pi^0]_D h^-) +  \Gamma(\Bp \to [h^\prime h^\prime \pi^0]_D h^+)  }.
\label{eq:aqglw}
\end{equation}

The relations between $R_{{\rm qGLW}}^{h^\prime h^\prime \pi^0}$ and $A_{{\rm qGLW}(K)}^{h^\prime h^\prime \pi^0}$, the most sensitive to $\gamma$ of the two asymmetries, and the underlying physics parameters are
\begin{eqnarray}
R_{{\rm qGLW}}^{h^\prime h^\prime \pi^0}  & = & 1 \, + \, (r_B)^2 + (2F^{h^\prime h^\prime \pi^0}_+ -1) 2r_B\cos\delta_B \cos \gamma,  \label{eq:rqglw_phy}\\
A_{{\rm qGLW}(K)}^{h^\prime h^\prime \pi^0} &=& (2F^{h^\prime h^\prime \pi^0}_+ -1 )  2r_B \sin\delta_B \sin\gamma /R_{{\rm qGLW}}^{h^\prime h^\prime \pi^0}\label{eq:aqglw_phy}.
\end{eqnarray} 
The small effects of  $D^0 \Dzb$ mixing are neglected, but can be accommodated if required~\cite{Nayak:2014tea}.
A recent analysis using CLEO-c data~\cite{Nayak:2014tea} has used decays of coherently produced $D \Db$ pairs to determine $F^{\pi^+ \pi^- \pi^0}_+ = 0.968 \pm 0.018$ and $F^{K^+K^-\pi^0}_+ = 0.731 \pm 0.062$.  The  high value of $F^{\pi^+ \pi^- \pi^0}_+$  implies that the decay $D^0 \to \pi^+ \pi^- \pi^0$
is very close to being a \CP-even eigenstate and the interference terms in Eqs.~\ref{eq:rqglw_phy} and~\ref{eq:aqglw_phy} suffer very little dilution, tending towards the equivalent GLW \CP-even expressions. 

When measuring \CP asymmetries at the LHC, it is necessary to allow for the possibility that the initial state may contain different numbers of \Bm and \Bp mesons.  Therefore, a production asymmetry,
\begin{equation}
A_{\rm Prod} \equiv \frac{\sigma(\Bm) - \sigma(\Bp)}{\sigma(\Bm) + \sigma(\Bp)},
\label{eq:aprod}
\end{equation}
is defined where $\sigma(\Bm)$ and $\sigma(\Bp)$ are the cross-sections for the production of \Bm and \Bp mesons, respectively, within the LHCb acceptance.

To summarise, twelve observables are measured in total: the two ADS asymmetries $A^{K\pi\pi^0}_{{\rm ADS} (h)}$,  two ratios $R^{K\pi\pi^0}_{{\rm ADS} (h)}$  and the asymmetry $A^{K\pi\pi^0}_K$; the four quasi-GLW asymmetries $A_{{\rm qGLW}(h)}^{h^\prime h^\prime \pi^0}$ and two ratios $R_{{\rm qGLW}}^{h^\prime h^\prime \pi^0}$; and the \Bp/\Bm production asymmetry, $A_{\rm Prod}$.

%% file: detector.tex
\section{The LHCb detector and data set}
\label{sec:Detector}

The analysis uses data collected by LHCb in $pp$ collisions at $\sqrt{s}=7$~TeV in 2011 and 8~TeV in 2012, corresponding to integrated luminosities of 1.0~\invfb and 2.0~\invfb, respectively.
The \lhcb detector~\cite{Alves:2008zz,LHCb-DP-2014-002} is a single-arm forward
spectrometer covering the \mbox{pseudorapidity} range $2<\eta <5$,
designed for the study of particles containing \bquark or \cquark
quarks. The detector includes a high-precision tracking system
consisting of a silicon-strip vertex detector surrounding the $pp$
interaction region, a large-area silicon-strip detector located
upstream of a dipole magnet with a bending power of about
$4{\rm\,Tm}$, and three stations of silicon-strip detectors and straw
drift tubes placed downstream of the magnet.
The polarity of the dipole magnet is reversed periodically throughout data-taking in order to combat systematic biases due to possible detector asymmetries.
The tracking system provides a measurement of momentum, \ptot, of charged particles with
a relative uncertainty that varies from 0.5\% at low momentum to 1.0\% at 200\gevc.
The minimum distance of a track to a primary vertex, the impact parameter, is measured with a resolution of $(15+29/\pt)\mum$,
where \pt is the component of the momentum transverse to the beam, in \gevc.
Different types of charged hadrons are distinguished using information
from two ring-imaging Cherenkov detectors. 
Photons, electrons and hadrons are identified by a calorimeter system consisting of
scintillating-pad and preshower detectors, an electromagnetic
calorimeter and a hadronic calorimeter. Muons are identified by a
system composed of alternating layers of iron and multiwire
proportional chambers.
The online event selection is performed by a trigger~\cite{LHCb-DP-2012-004}, 
which consists of a hardware stage, based on information from the calorimeter and muon
systems, followed by a software stage, which applies a full event
reconstruction.  Offline a loose selection based on a decision tree 
algorithm~\cite{BBDT} is run to reduce the size of the sample prior to 
final analysis.  

Approximately one million simulated events (after geometric detector acceptance) of each class of signal decay are used in the analysis, as well as a large inclusive sample of generic $B_q \to DX$ decays, where $q\in\{u,d,s\}$.
In the simulation, $pp$ collisions are generated using
\pythia~\cite{Sjostrand:2006za,*Sjostrand:2007gs} 
 with a specific \lhcb
configuration~\cite{LHCb-PROC-2010-056}.  Decays of hadronic particles
are described by \evtgen~\cite{Lange:2001uf}, in which final-state
radiation is generated using \photos~\cite{Golonka:2005pn}. The
interaction of the generated particles with the detector, and its response,
are implemented using the \geant
toolkit~\cite{Allison:2006ve, *Agostinelli:2002hh} as described in
Ref.~\cite{LHCb-PROC-2011-006}.

%% file: selection.tex
\section{Candidate selection}
\label{sec:Selection}

The events used in the analysis must be selected by the hardware trigger, either for the case where the \Bmp candidate triggered the event via the hadronic calorimeter (and not the muon system), or where the event was triggered independently of the \Bmp candidate.  The study is performed with \BToDh candidates, where the \D meson decays to a three-body final state composed of any combination of two charged kaons and pions and a \piz candidate.  The \piz is identified by a decay to two photons, as recorded by the electromagnetic calorimeter.

All candidates passing the \BToDh reconstruction are required to have an invariant mass in the range of 5080--5900 \mevcc.  The mass of the reconstructed \D candidate is required to be within $\pm50$ \mevcc of the nominal \Dz mass~\cite{PDG2014}.  In addition, the mass of the \piz candidate must be within $\pm20$ \mevcc of the nominal \piz mass~\cite{PDG2014}.  Both of these mass windows correspond to approximately plus or minus twice the mass resolution of the respective reconstructed particles.  The \piz candidate must also have a momentum of $\pt >0.5$ \gevc and $p >1.0$ \gevc.  The bachelor particle is required to satisfy $0.5 < \pt < 10$ \gevc and $5 < p < 100$ \gevc, while the charged \D daughters must have $\pt > 0.25$ \gevc.  In order to improve the resolution of the mass of the \Bmp candidate, the decay chain is refitted \cite{Hulsbergen:2005pu} constraining the positions of the \Bmp and \D vertices, while at the same time constraining the \D candidate to its nominal mass.

In addition to these selection criteria, further background suppression is achieved through the use of a boosted decision tree (BDT) discriminator \cite{Breiman} using the \textit{GradientBoost} algorithm \cite{Friedman2002367}.  The BDT is trained using a signal sample of \BToDh events from simulation and a sample of pure combinatorial background from data with \Bmp candidates' invariant mass greater than 5900 \mevcc, which are not used in the invariant mass fit.  The BDT utilises a variety of properties associated to each signal candidate.  These properties include: $p$ and \pt of the \D meson, the \D daughter candidates and the bachelor particle; and the \chisqip of the \D meson, charged \D daughter candidates, bachelor particle and the \Bmp meson (where \chisqip is defined as the difference between the \chisq of the primary vertex (PV) reconstructed with and without the particle of interest).  Other properties include: the flight distance from the PV for the \Bmp and \D candidates; vertex quality, \chisq per degree of freedom, for the \Bmp and \D candidates; and the angle between the line connecting the PV to the particle's decay vertex and the particle's momentum vector for the \Bmp and \D candidates.  Another characteristic used in the BDT is an isolation variable representative of the \pt imbalance surrounding a \Bmp candidate.  The variable is defined as
\begin{equation}
	A_{\pt}  = \frac{\pt(\Bmp) - \sum_n \pt}{\pt(\Bmp) + \sum_n \pt},
\end{equation}
where the sum is performed over the $n$ tracks lying within a cone around the candidate, excluding the tracks related to the signal.  The cone is defined by a circle of radius 1.5 units in the plane of pseudorapidity and azimuthal angle (measured in radians).  No PID information is used as an input variable; consequently the BDT has similar performance for both the \BToDK and \BToDPi decay modes, with some slight variation arising due to differences in kinematics between the two.

The optimal cut value of the BDT is determined by optimising the metric $s/\sqrt{s+b}$, where $s$ is the expected signal yield in the suppressed \BToDK ADS mode and $b$ is the combinatoric background level as taken from the favoured mode, which is expected to have comparable background levels to the suppressed mode.  The expected signal yield is calculated as the yield in the favoured \BToDPi ADS mode scaled by the predicted branching fraction of the \BToDK mode and by the expected ratio between the suppressed and favoured ADS modes, while taking into account differences in PID efficiency.  Assessment of this $s/\sqrt{s+b}$ metric finds a working point where a signal efficiency of ~85\% is expected while rejecting $>99\%$ of combinatorial background.  A similar optimisation procedure performed using the \BToDKWithDToPiPiPi0 and \BToDKWithDToKKPi0 decays returns a comparable working point, and thus the same requirement is imposed in the selection of the quasi-GLW modes, as well as the ADS modes.

Particle identification, essential for the distinction between \BToDK and \BToDPi candidates, is quantified by differences between the logarithm of likelihoods, $\ln \mathcal{L}_h$, under five separate mass hypotheses, $h \in \{e, \mu, \pi, K, p\}$ (DLL).  For the daughters from the \D candidate, the kaon must satisfy $\text{DLL}_{K\pi} \equiv \ln \mathcal{L}_K - \ln \mathcal{L}_\pi >2$, while the charged pion is required to satisfy $\text{DLL}_{K\pi}<-2$.  Candidates with a bachelor having $\text{DLL}_{K\pi}>4 $ are selected into the \BToDK sample (they are said to have \textit{passed} the PID requirement) while those that do not are placed in the \BToDPi sample (they are said to have \textit{failed} the PID requirement).

Additional restrictions are imposed after the BDT and the PID requirements in order to remove specific sources of background.  Contributions from genuine \Bmp decays that do not include a \D meson are suppressed through a selection requirement on the flight distance significance, ${\rm FD}_D$, defined as the distance between the \Bmp and \D candidate vertices, divided by the uncertainty on this measurement.  A requirement of ${\rm FD}_D > 2$ is applied.  The total branching fractions of \Bmp to four-body charmless states with a \piz are currently unmeasured and their contribution is estimated by studying the contamination of three-body charmless modes to the $B^\mp \to [K^\mp\pi^\pm, \pi^\mp K^\pm]_D h^\mp$ spectra and scaling it according to the known branching fractions.  The efficiency of the ${\rm FD}_D$ requirement is evaluated using simulated \bquark-hadron decays to four-body charmless states with a neutral pion.  The requirement is found to be $93\%$ effective in the suppression of this background, a value compatible with that seen in data for the three-body charmless states.  From these studies, it is determined that the charmless backgrounds contribute $4\pm1$, $1\pm1$, $4\pm1$ and $3\pm1$ candidates to the summed-by-charge selections of \favBToDHWithDToKPiPi0 , \supBToDHWithDToKPiPi0 , \BToDHWithDToPiPiPi0 and \BToDHWithDToKKPi0 , respectively.

The suppressed \supBToDHWithDToKPiPi0 decays are subject to potential contamination from \BToDHWithDToPiPiPi0 and \BToDHWithDToKKPi0 decays where one of the charged pions or kaons from the \D candidate is misidentified as a charged kaon or pion, respectively.  Studies performed using simulated events demonstrate that such contamination is minimal, contributing $1 \pm 1$ candidate to each \supBToDHWithDToKPiPi0 decay mode.  Similarly, there is potential cross-feed from favoured \favBToDHWithDToKPiPi0 decays in the suppressed ADS signal samples in which a \Kpm and \pimp are doubly misidentified as a \pipm and a \Kmp, respectively.  This contamination is reduced by vetoing any suppressed candidate whose reconstructed \D mass, under the exchange of mass hypotheses between the daughter kaon and charged pion, lies within $\pm 30$ \mevcc of the nominal \D mass.  Study of the cross-feed contamination in the mass sidebands of the \D candidates allows for an estimate of the residual contamination in the signal region.  After all selection requirements, this residual cross-feed is estimated to be $(3.1 \pm 0.2) \times 10^{-4}$ of the total favoured \favBToDHWithDToKPiPi0 events.

For each event, only one candidate is selected for analysis.  In the 3.8\% of cases where more than one candidate is present in an event, a choice is made by selecting the candidate with the \Bmp decay vertex with the smallest \chisq per degree of freedom.

%% file: fit.tex
\section{Invariant mass fit}
\label{sec:Fit}

The observables of interest are determined with a binned maximum-likelihood fit to the invariant mass of the selected \Bmp candidates.  A total of sixteen disjoint subsamples (the favoured and suppressed ADS modes and the two quasi-GLW modes, separated according to the charge of the bachelor meson, and by the bachelor PID requirement) are fitted simultaneously.  The total probability density function (PDF) used in the fit is built from five main sources, described below, representing different categories of candidates in each subsample.

The \BToDPi signal events are modelled through the use of a modified Gaussian function,
	\begin{equation}
		f(m) \propto \exp \left( \frac{-(m-\mu)^2}{2\sigma^2 + (m-\mu)^2 \alpha_{L,R}} \right).
		\label{eq:cruijff}
	\end{equation}
This expression describes an asymmetric peak of mean $\mu$ and width $\sigma$ where the values of $\alpha_L (m < \mu)$ and $\alpha_R (m > \mu)$ parameterise the tails of the distribution to the left and to the right of the peak, respectively.  These signal events originate from subsamples that fail the bachelor PID requirement for charged kaons.  Genuine \BToDPi candidates that pass the PID requirement are reconstructed as \BToDK.  Since these candidates are reconstructed under an incorrect mass hypothesis, they represent a displaced mass peak with a tail that extends to higher invariant mass.  Such misidentified candidates are modelled by the sum of two Gaussian functions, modified to include tail components similar to that of Eq.~\ref{eq:cruijff}.  The two modified Gaussian functions share a mean, but have two separate width parameters that are permitted to float.  For the signal peaks, all of the parameters are permitted to vary, with the exception of the lower-mass tail, which is fixed to the value found in simulation, to ensure fit stability, and later considered as a source of systematic uncertainty.  The same shape is used for \Bm and \Bp decays, although the means are allowed to be different.  In addition, while the \favBToDHWithDToKPiPi0 and \supBToDHWithDToKPiPi0 signal shapes share the same width, this parameter is permitted to vary for the \BToDHWithDToKKPi0 and \BToDHWithDToPiPiPi0 modes.

The \BToDK signal events, from the subsamples that pass the PID requirement on the bachelor, are modelled using the same modified Gaussian function of Eq.~\ref{eq:cruijff}.  All of the shape parameters are identical to those of the \BToDPi modes, except for the width, which is fixed at $(95 \pm 2)\%$ of that of the \BToDPi modes, based upon studies made using simulated events.  Genuine \BToDK candidates that fail the PID selection (and thus represent misidentified \BToDPi events) are described using a fixed shape from simulation that is later varied to assign a systematic uncertainty.

Partially reconstructed $b$-hadron decays are found in the invariant mass region below the \Bmp mass.  However, a portion may enter the signal region.  Of particular concern are \Bmp ($\BorBbar^0$) decays involving a neutral (charged) \Dstar meson, where the \Dstar decays to a $\DorDbar^0$ and a neutral (charged) pion with this latter particle missed in reconstruction, leading to the same final state as in the channels of interest.  The $\DorDbar^{*0}$ may also decay via the $\DorDbar^{*0} \to \DorDbar^0 \g$ channel.  When the \g is missed in reconstruction, such decays may also mimic the desired signal candidates.  There are also further contributions from \Bmp~($\BorBbar^0$) decays to $\DorDbar^0$ and a neutral (charged) $\rho$ or \Kstar, where the vector meson decays into an $h^\pm \pi^\mp$ ($h^\pm \piz$) state from which the $\pi^\mp$ (\piz) is missed in reconstruction.  These partially reconstructed decays are described by parabolic functions representative of the decays in question, that have been convolved with a double Gaussian to account for detector resolution.  The yields of these background components vary independently in the fit, with no assumption of \CP symmetry.  Additionally, partially reconstructed $\BsorBsbar^0 \to \D \Kmp \pipm$ decays and their charge-conjugated modes are considered as background sources to the ADS \BToDh modes.  PDFs for this background are determined from simulation and fixed in the invariant mass fit.  The $\BsorBsbar^0$ yields are permitted to float, but \CP symmetry is assumed given the limited interference effects due to Cabibbo suppression.

Wrongly reconstructed \D meson decays are a source of background under the signal peaks.  These are primarily decays where the \piz candidate is not a daughter of the \D meson, but is wrongly assigned as such.  In the final fits, these contributions are modelled using a modified Gaussian function with a tail parameter, where this component and the width are permitted to vary, but the mean is fixed based on a study in data.  In this study, a binned-maximum likelihood fit is performed to the \D mass distribution in a region of $\pm250$ \mevcc about the nominal \D mass~\cite{PDG2014}, where the signal and background contributions are modelled separately.  The sPlot method~\cite{Pivk:2004ty} is used to assign signal and background weights to the candidates and the \Bmp invariant mass distribution is then plotted using the background weights in order to ascertain how the wrongly reconstructed \D background contribution distributes itself in the \Bmp mass spectrum.  This study indicates that the background can be described by using the function of Eq.~\ref{eq:cruijff} with a single tail parameter.  As such, in the final fits, the wrongly reconstructed \D meson background is modelled as a fully floating modified Gaussian function, except for the mean that is fixed.  The value of the fixed parameter is varied in order to assess a systematic uncertainty.

A linear approximation is adequate to describe the distribution of combinatorial background across the relevant invariant mass spectrum.  All \BToDK modes and all  \BToDPi modes share the same respective shapes, though yields vary independently.  This allows for greater fit stability as the low statistics modes share fit information from the higher statistics modes.

The measured signal yields allow the fit to determine the observables of interest.  For example, the relationship between $n^{DK+}_{\pi K \piz}$, the yield of the decay $\Bp \to [\pip \Km \piz]_D \Kp$, and the physics observables is given by
\begin{equation}
	n^{DK+}_{\pi K \piz} = \frac{n^{D\pi}_{K \pi \piz} \cdot R^{K\pi\piz}_{K/\pi} \cdot R^{K\pi\piz}_{{\rm ADS}(K)} \cdot \epsilon_{\rm exp}}
	{1 + \left[\frac{1+A^{K\pi\piz}_{{\rm ADS}(K)}}{1-A^{K\pi\piz}_{{\rm ADS}(K)}} \cdot \frac{1+A_{\rm Prod}}{1-A_{\rm Prod}} \cdot \frac{1+A_{\rm det}}{1-A_{\rm det}}   \right]},
\label{eq:sup_yields}
\end{equation}
where $\epsilon_{\rm exp}$ represents experimental selection efficiency effects and $A_{\rm det}$ are detector-related asymmetries (both of these are further discussed in Sect.~\ref{sec:Results}) and $n^{D\pi}_{K\pi\piz}$ is the total yield of \favBToDPiWithDToKPiPi0 decays.  In the fit, an analogous expression to Eq.~\ref{eq:sup_yields} is used for the corresponding \Bm decay as well as comparable equations for the other decay modes and their associated \CP observables.

\begin{figure}[htbp]
	\centering
	\includegraphics[scale=0.75]{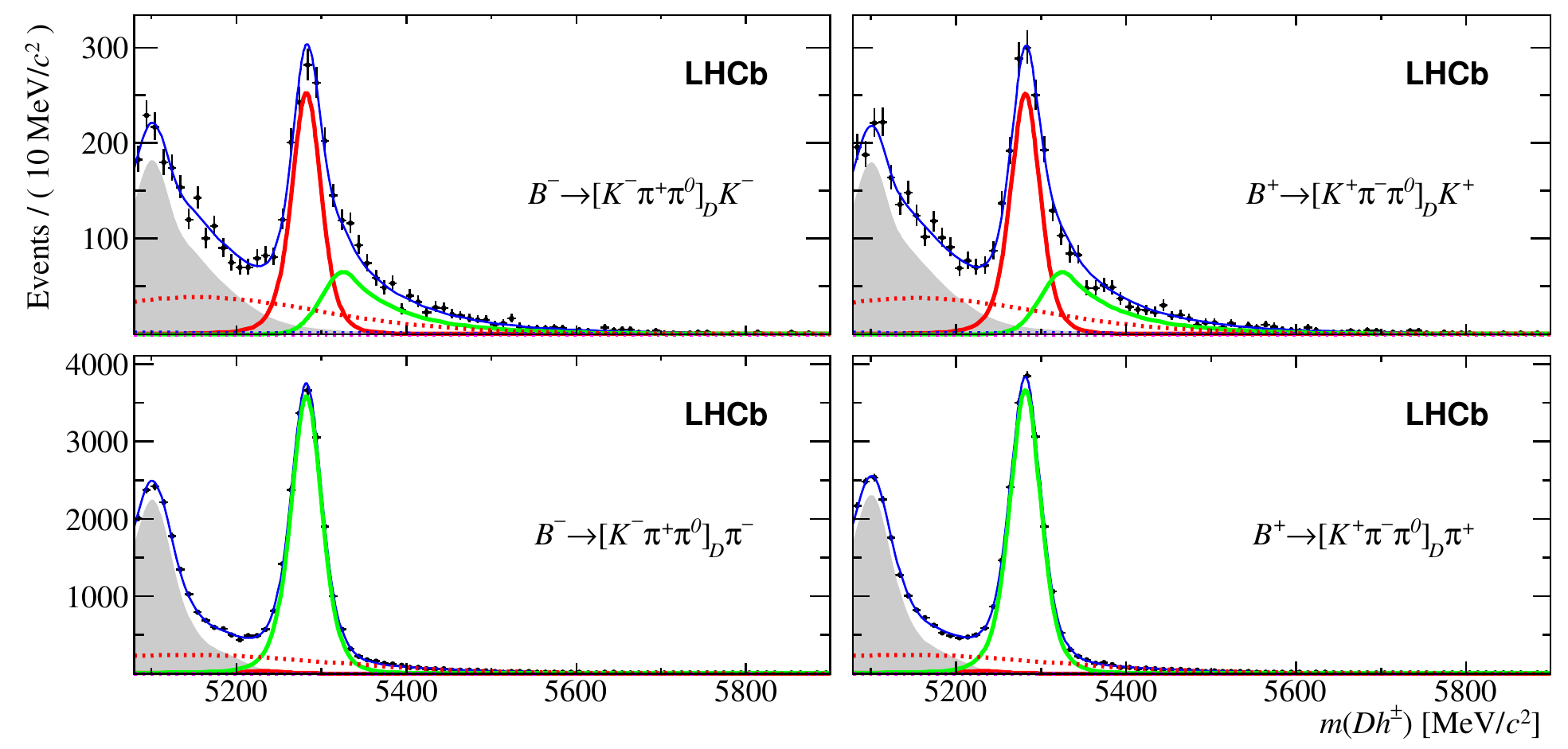}
	\caption{Invariant mass distributions of selected \favBToDHWithDToKPiPi0 candidates, separated by \B hadron charge.  \BToDK signal events are in the upper plots and \BToDPi events are in the lower plots.  The solid dark (red) curve represents \BToDK events and the solid light (green) curve represents \BToDPi events.  The solid (grey) shape indicates partially reconstructed \Bmp decays and the heavy dotted (red) curve indicates wrongly reconstructed \D decays.  The solid (blue) line represents the total PDF and includes the combinatorial component.}
	\label{fig:fav_kpipi0}
\end{figure}

\begin{figure}[htbp]
	\centering
	\includegraphics[scale=0.75]{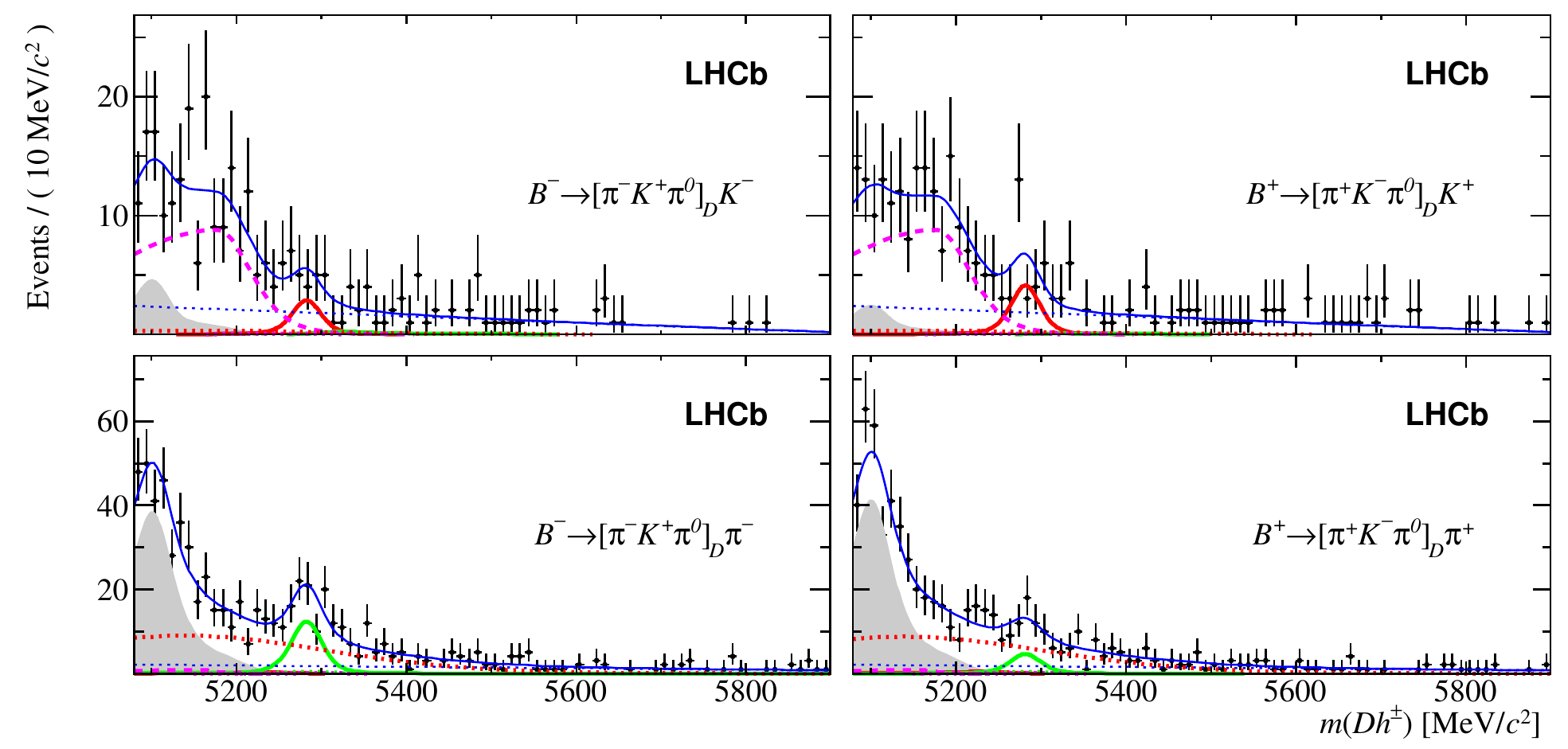}
	\caption{Invariant mass distributions of selected \supBToDHWithDToKPiPi0 candidates.  See the caption of Fig.~\ref{fig:fav_kpipi0} for a full description.  The lightly dotted (blue) line represents the combinatorial component and the long-dashed (magenta) line indicates contributions from partially reconstructed $\Bs \rightarrow D \Kmp \pipm$ decays where the pion is not reconstructed.} 
	\label{fig:sup_kpipi0}
\end{figure}

\begin{figure}[htbp]
	\centering
	\includegraphics[scale=0.75]{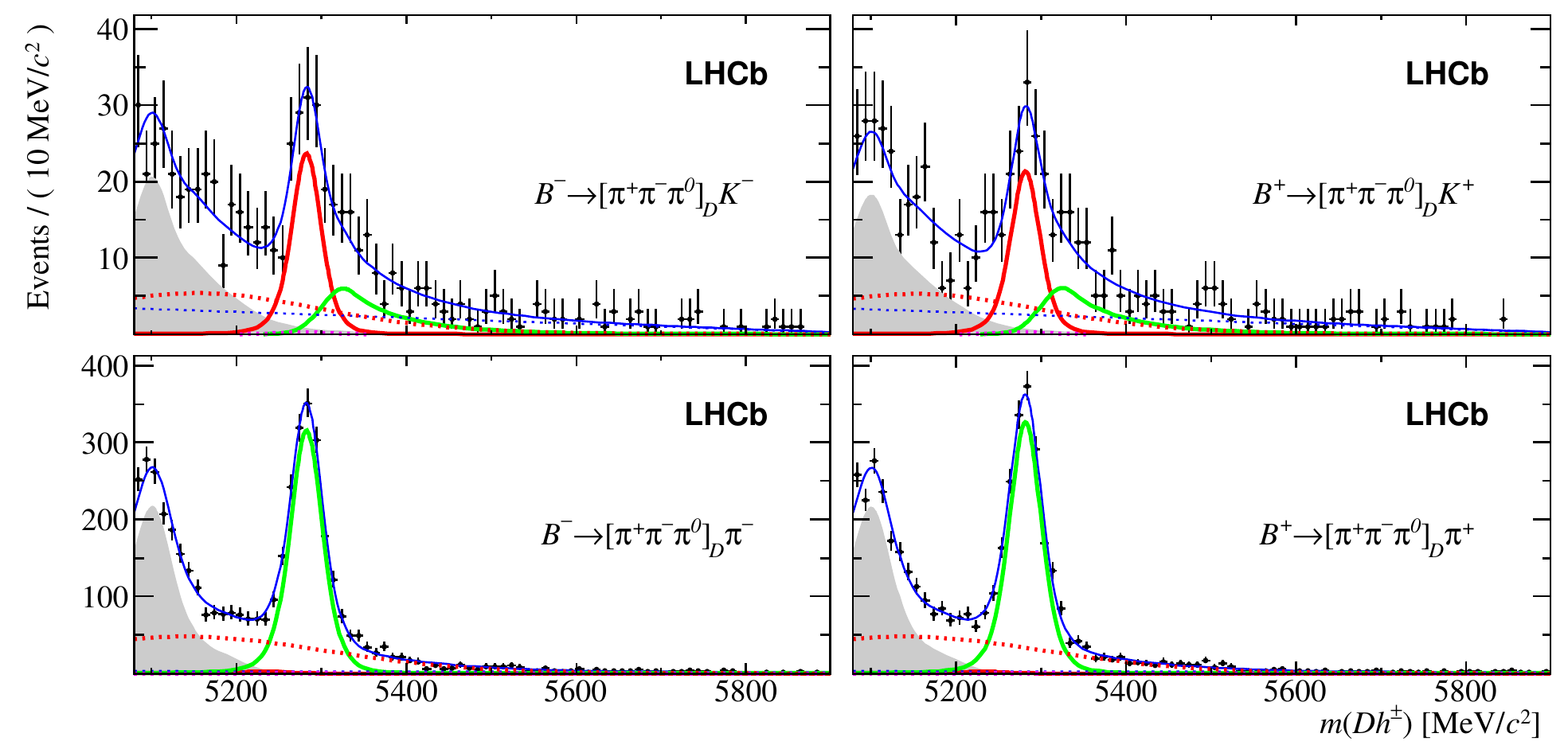}
	\caption{Invariant mass distributions of selected \BToDHWithDToPiPiPi0 candidates.  See the caption of Fig.~\ref{fig:fav_kpipi0} for a full description.  The lightly dotted (blue) line represents the combinatorial component.}
	\label{fig:pipipi0}
\end{figure}

\begin{figure}[htbp]
	\centering
	\includegraphics[scale=0.75]{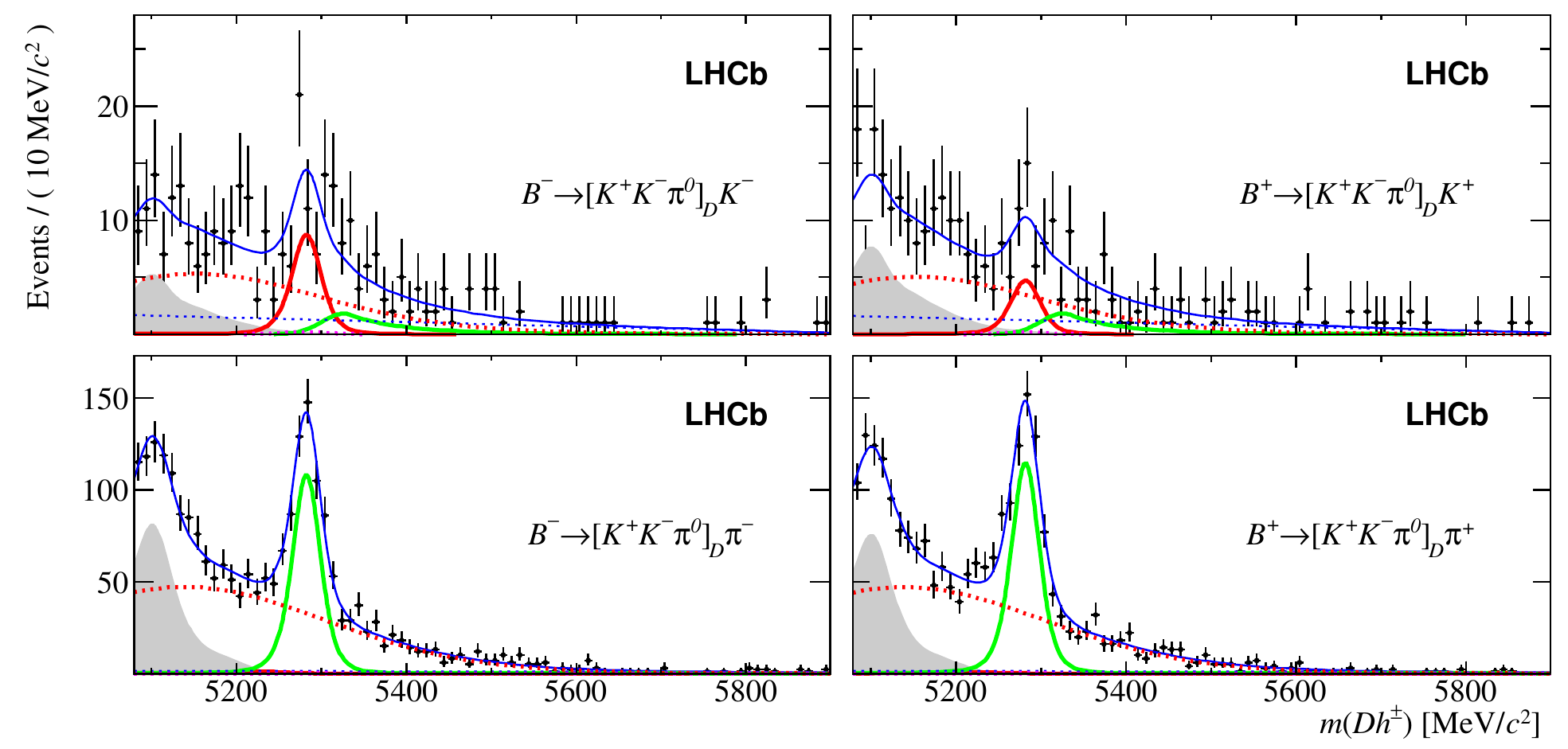}
	\caption{Invariant mass distributions of selected \BToDHWithDToKKPi0 candidates.  See the caption of Fig.~\ref{fig:fav_kpipi0} for a full description.  The lightly dotted (blue) line represents the combinatorial component.}
	\label{fig:kkpi0}
\end{figure}

The fit is performed such that all of the observables defined by Eqs.~\ref{eq:rads},~\ref{eq:aads},~\ref{eq:afavoured},~\ref{eq:rkpi1}, \ref{eq:aqglw} and~\ref{eq:aprod} are free parameters.  The signal yields for the decay modes of interest are presented in Table~\ref{tab:yields}.  The uncertainties are statistical only; the systematic uncertainties are discussed in Sect.~\ref{sec:Results}.  The corresponding invariant mass spectra, separated by the charge of the \Bmp candidate, are presented in Figs.~\ref{fig:fav_kpipi0}, \ref{fig:sup_kpipi0}, \ref{fig:pipipi0} and \ref{fig:kkpi0}.

\begin{table}[htbp]
	\centering
	\caption{The final signal yields, split in categories based on the charges of the \B hadron (only statistical uncertainties are shown).}
	\begin{tabular}{lrclr}
	\cline{1-2} \cline{4-5}
	\Bm decay channel & Yield & & \Bp decay channel & Yield \\ \cline{1-2} \cline{4-5}
	$\Bm \to [\Km \pip \piz]_D\pim$ & $18\,854 \pm 176$ & ~~~~&$\Bp \to [\Kp \pim \piz]_D\pip$ & $18\,882 \pm 176$ \\
	$\Bm \to [\Km \pip \piz]_D\Km$  & \phantom{18}$1\,478 \pm \phantom{4}39$ &  ~~~~&$\Bp \to [\Kp \pim \piz]_D\Kp$  & \phantom{18}$1\,442 \pm \phantom{4}39$ \\
	$\Bm \to [\pim \Kp \piz]_D\pim$ & $\phantom{188}63 \pm \phantom{4}13$ & ~~~~&$\Bp \to [\pip \Km \piz]_D\pip$ & $\phantom{188}25 \pm \phantom{4}13$ \\
	$\Bm \to [\pim \Kp \piz]_D\Km$ & $\phantom{188}16 \pm \phantom{44}9$ & ~~~~&$\Bp \to [\pip \Km \piz]_D\Kp$ & $\phantom{188}24 \pm \phantom{44}9$ \\
	$\Bm \to [\pip \pim \piz]_D\pim$ & $1\,716 \pm \phantom{4}55$ & ~~~~ & $\Bp \to [\pip \pim \piz]_D\pip$ & $1\,772 \pm \phantom{4}55$ \\
	$\Bm \to [\pip \pim \piz]_D\Km$ & $\phantom{18}139 \pm \phantom{4}19$ & ~~~~ & $\Bp \to [\pip \pim \piz]_D\Kp$ & $\phantom{18}125 \pm \phantom{4}19$ \\
	$\Bm \to [\Kp \Km \piz]_D\pim$ & \phantom{189}$509 \pm \phantom{4}34$ & ~~~~ & $\Bp \to [\Kp \Km \piz]_D\pip$ & \phantom{189}$541 \pm \phantom{4}34$ \\
	$\Bm \to [\Kp \Km \piz]_D\Km$ & $\phantom{188}49 \pm \phantom{4}12$ & ~~~~ & $\Bp \to [\Kp \Km \piz]_D\Kp$ & $\phantom{188}27 \pm \phantom{4}12$ \\
\cline{1-2} \cline{4-5}
	\end{tabular}
	\label{tab:yields}
\end{table}

%% file: results.tex
\section{Systematic uncertainties and results}
\label{sec:Results}

In addition to the sources of systematic uncertainties originating from fixed PDF parameters in the fit, there are several other sources that are considered.  In the favoured and suppressed ADS modes, the ratio $R^{K\pi\piz}_{K/\pi}$ is fixed at 7.74\% based on the measurement performed in Ref.~\cite{LHCb-PAPER-2012-001} and is assigned a systematic uncertainty of 0.22\%, as per the uncertainties of that analysis.  The \BToDK versus \BToDPi ratio, however, is permitted to vary in the \BToDHWithDToPiPiPi0 and \BToDHWithDToKKPi0 analyses as it must be measured for each mode in order to determine the $R^{h'h'\pi^0}_{\rm qGLW}$ observables.

The proportion of \BToDh samples passing or failing the PID requirements is determined from a sample of more than 100 million $\Dstarpm$ decays reconstructed as $\Dstarpm \to \D \pipm, \D \to \Kmp \pipm$.  This reconstruction is performed entirely using kinematic variables and provides a high-purity calibration sample of \Kpm and \pipm tracks.  The PID efficiency varies as a function of track momentum, pseudorapidity and detector occupancy \cite{LHCb-DP-2012-003}.  The average PID efficiency of the signal is determined by reweighting the calibration spectra in these variables to those of the candidates in the favoured ADS sample.  This average PID efficiency is evaluated to be 84.5\% and 96.3\% for kaons and pions, respectively.  Systematic uncertainties of 0.5\% and 0.8\% for bachelor pions and bachelor kaons, respectively, are assigned to the efficiencies, which arise from the reweighting procedure.

Due to differences in interactions with the detector material, a small negative asymmetry is expected in the detection of \Km and \Kp mesons.  An asymmetry for pions may also be present and is assigned a value of $(0.0 \pm 0.3)\%$ \cite{LHCb-PAPER-2012-052}.  The difference between the kaon and pion detection asymmetries is taken to be $-(1.1 \pm 0.4)\%$ from studies performed in Ref.~\cite{LHCb-PAPER-2014-013}.  These asymmetry values also account for the physical asymmetry of the left and right sides of the detector, after summing the data sets from both magnet polarities.  There is no systematic uncertainty associated with the possible difference in number of \Bm and \Bp mesons, since the production asymmetry $A_{\rm Prod}$ is a variable parameter in the fit.

The measured observables in the analysis are related to the ratio of relative efficiencies between the \BToDK and \BToDPi modes, $\epsilon_{\B \to DK} / \epsilon_{\B \to D\pi}$, independent of PID effects.  These ratios relate the efficiency differences due to trigger, reconstruction and selection effects.  They are measured in simulation to be $(97.5 \pm 3.4)\%$ for the \favBToDHWithDToKPiPi0 and \supBToDHWithDToKPiPi0 modes, $(95.7 \pm 2.8)\%$ for the \BToDHWithDToPiPiPi0 modes and $(98.9 \pm 2.8)\%$ for the \BToDHWithDToKKPi0 modes.  The uncertainties listed are based on the finite size of the simulated samples and account for the imperfect modelling of pion and kaon absorption rates in the detector material.

In order to estimate the systematic uncertainties from the sources described in this section and in Sect.~\ref{sec:Fit}, the fit is performed many times, varying each source by its assigned uncertainty, under the assumption that the uncertainty is Gaussian distributed.  The spread (RMS) in the distribution of the fitted value of the observables is taken as the systematic uncertainty.  These uncertainties are summarised in Table~\ref{tab:systematics}.

\begin{table}[!tb]
\centering
\caption{Systematic uncertainties on the observables, multiplied by a factor of $10^{3}$.  `PID' refers to the fixed PID efficiency attributed to the bachelor tracks.  `PDFs' refers to the uncertainties based on fixed parameters in the PDF shapes that are used in the invariant mass fit.  `Sim' refers to the use of simulation to calculate relative efficiencies between the \BToDK and \BToDPi modes, in addition to the estimated charmless background contributions and the fixed $DK$ to $D\pi$ ratio on the ADS modes.  `$A_{\rm instr}$' refers to the interaction and detection asymmetries.  The `Total' column represents the sum in quadrature of all of the categories of systematic uncertainties.}

\begin{tabular}{l r r r r r}
\hline
& PID & PDFs & Sim & $A_{\rm instr}$ & Total \\
\hline
$A_{{\rm ADS}(K)}^{K \pi \piz}$ & 3.4 & 39.6 & 8.7 & 5.7 & 41.1 \\
$A_{{\rm ADS}(\pi)}^{K\pi\piz}$ & 1.6 & 7.5 & 4.5 & 6.9 & 11.3 \\
$A_{{\rm qGLW}(K)}^{KK\piz}$ & 5.1 & 10.2 & 18.8 & 2.1 & 22.1 \\
$A_{{\rm qGLW}(K)}^{\pi\pi\piz}$ & 0.9 & 7.9 & 7.3 & 0.9 & 10.8 \\
$A_{{\rm qGLW}(\pi)}^{KK\piz}$ & 0.8 & 2.2 & 1.2 & 4.4 & 5.1 \\
$A_{{\rm qGLW}(\pi)}^{\pi\pi\piz}$ & 0.3 & 0.9 & 0.7 & 4.2 & 4.4 \\
$A_{K}^{K\pi\piz}$ & 0.4 & 0.9 & 1.4 & 4.2 & 4.6 \\
$R_{{\rm ADS}(K)}^{K\pi\piz}$ & 0.3 & 2.0 & 0.6 & 0.1 & 2.1 \\
$R_{{\rm ADS}(\pi)}^{K\pi\piz}$ & 0.02 & 0.05 & 0.02 & 0.01 & 0.06 \\
$R^{KK\piz}_{{\rm qGLW}}$ & 23.8 & 24.9 & 36.5 & 7.7 & 50.8 \\
$R^{\pi\pi\piz}_{{\rm qGLW}}$ & 8.1 & 20.7 & 42.5 & 5.3 & 48.3 \\
$A_{\rm Prod}$ & 0.3 & 0.3 & 0.5 & 5.0 & 5.0 \\
\hline
\end{tabular}
\label{tab:systematics}
\end{table}

The values for the coherence factor, average strong-phase differences and \CP-even fraction reported in Refs.~\cite{Libby:2014rea} and \cite{Nayak:2014tea} assume a uniform acceptance across the three-body phase space of the \D decay, which is not the case in this analysis.  Studies are performed with amplitude models for the decays of interest and a modelling of the acceptance function derived from simulation to assess the impact upon these parameters arising from this source.  It is found that in all cases the biases are negligible compared to the assigned uncertainties.

The results for the observables, as determined by the fit, are
\begin{align*}
A^{K\pi \piz}_{{\rm ADS}(K)} &= -0.20\pm0.27\pm0.04 &\\
A^{K\pi \piz}_{{\rm ADS}(\pi)} &= 0.438\pm0.190\pm0.011 \\
A_{{\rm qGLW}(K)}^{KK\piz} &= 0.30 \pm 0.20 \pm 0.02 \\
A_{{\rm qGLW}(K)}^{\pi\pi\piz} &= 0.054 \pm 0.091 \pm 0.011\\
A_{{\rm qGLW}(\pi)}^{KK\piz} &= -0.030 \pm 0.040 \pm 0.005 \\
A_{{\rm qGLW}(\pi)}^{\pi\pi\piz} &= -0.016 \pm 0.020 \pm 0.004 \\
A^{K\pi\piz}_K &= 0.010 \pm 0.026 \pm 0.005\\
R^{K\pi \piz}_{{\rm ADS}(K)} &= 0.0140 \pm 0.0047 \pm 0.0021 \\
R^{K\pi \piz}_{{\rm ADS}(\pi)} &= 0.00235 \pm 0.00049 \pm 0.00006 \\
R_{\rm qGLW}^{KK\piz} &= 0.95  \pm 0.22 \pm 0.05 \\
R_{\rm qGLW}^{\pi\pi\piz} &= 0.98 \pm 0.11 \pm 0.05 \\
A_{\rm Prod} &= -0.0008 \pm 0.0055 \pm 0.0050 \\
\end{align*} 
where the first uncertainties are statistical and the second are systematic.

None of the asymmetry observables exhibit any significant \CP violation.  The results for the ADS observables are more precise than those obtained by previous experiments~\cite{Lees:2011up,Nayak:2013tgg} and are compatible with them.  Furthermore, apart from  $A^{\pi\pi\piz}_{{\rm qGLW}(K)}$, this is the first time that the quasi-GLW observables have been measured.
 
A likelihood-ratio test is used to assess the significance of the suppressed ADS signal yields, as well as those of the \BToDHWithDToKKPi0 decays.  This is performed by calculating the quantity $\sqrt{-2 \ln (\mathcal{L}_{\rm b} / \mathcal{L}_{\rm s+b})}$ where $\mathcal{L}_{\rm b}$ and $\mathcal{L}_{\rm s+b}$ are the maximum likelihood values of the background-only and signal-plus-background hypotheses, respectively.  Including systematic uncertainties, significances of $5.3\sigma$ and $2.8\sigma$ are found for the \supBToDPiWithDToKPiPi0 and \supBToDKWithDToKPiPi0 decays, respectively.  For the \BToDHWithDToKKPi0 selections, the \BToDPi mode is found to have a significance greater than $10\sigma$, while a significance of $4.5\sigma$ is measured for the \BToDK decay.

%% file: interpretation.tex
\section{Interpretation and conclusions}
\label{sec:Interpretation}

\begin{figure}[htbp]
	\centering
	\includegraphics[scale=0.45]{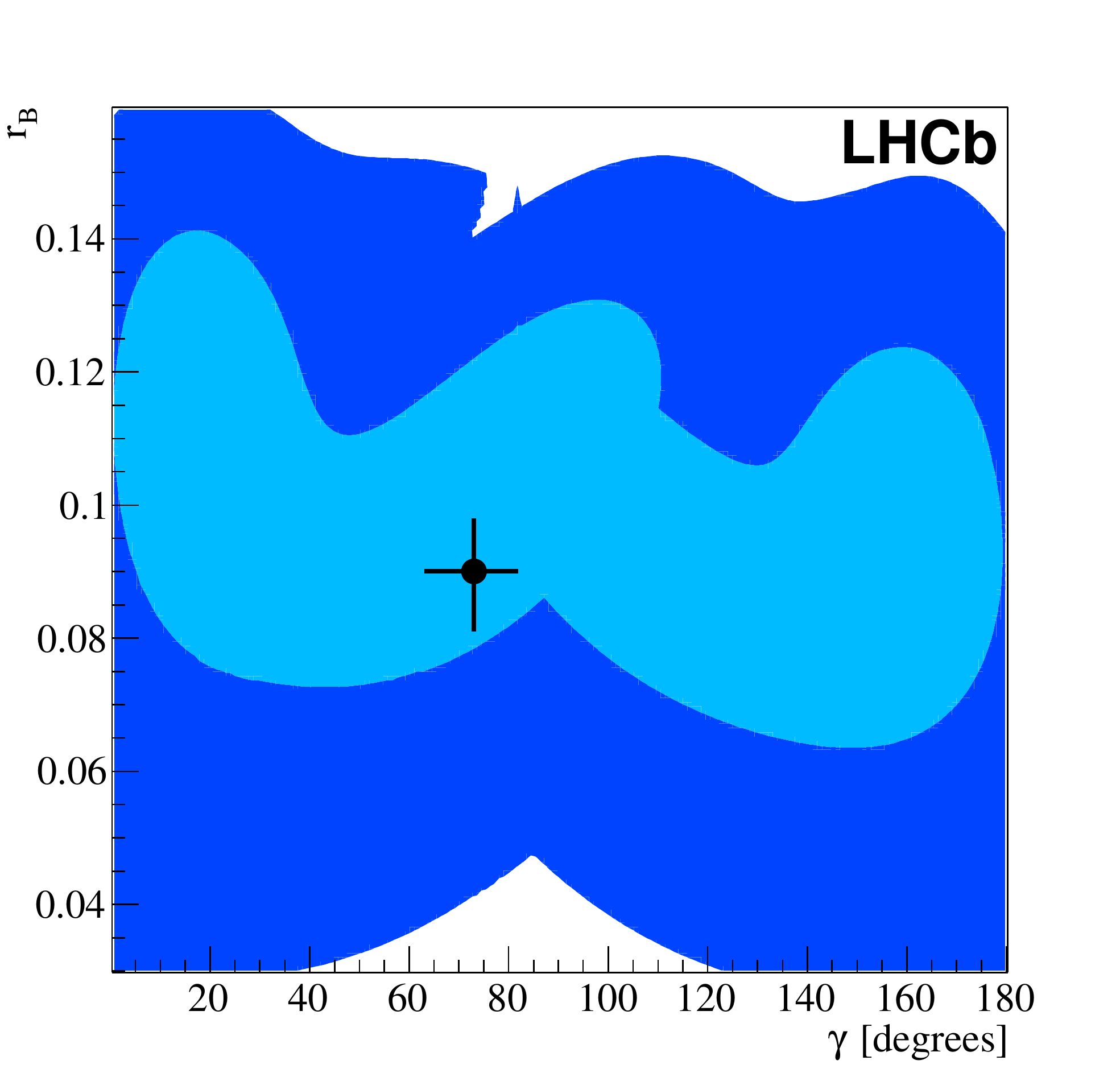}
	\caption{Scan of the \chisq probabilities over the \g--$r_B$ parameter space.  Shown are the $n\sigma$ profile likelihood contours, where $\Delta\chisq  = n^2$, with $n=1$ being the light (blue) shaded region, $n=2$ the dark (blue) shaded region and $n=3$ corresponding to the white area.  The result is seen to be compatible with the current LHCb measurement of \g and $r_B$, indicated by the point with error bars.
}
	\label{fig:rb_vs_gamma}
\end{figure}

\begin{figure}[htbp]
	\centering
	\includegraphics[scale=0.45]{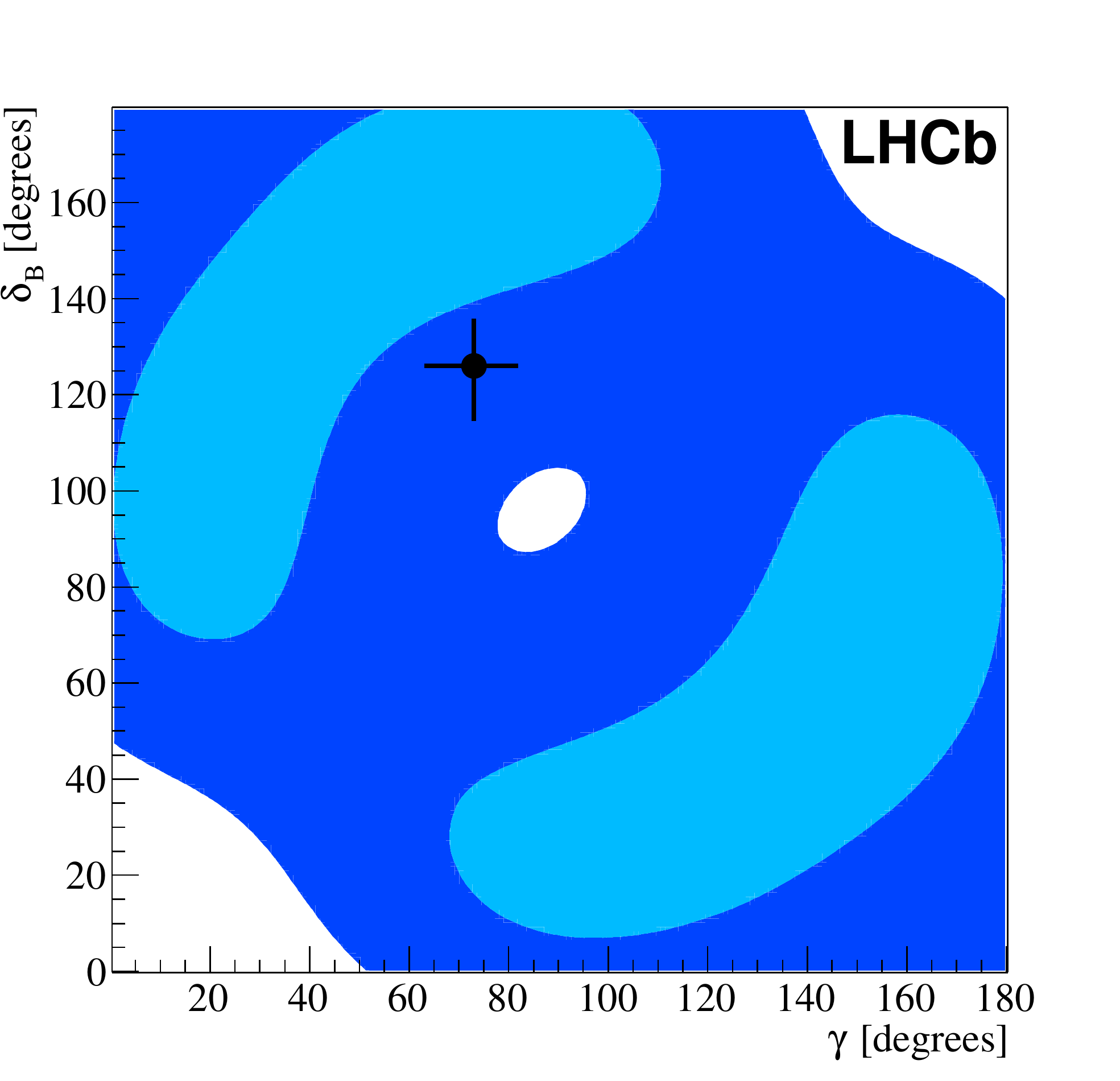}
	\caption{Scan of the \chisq probabilities over the \g--$\delta_B$ parameter space.  Shown are the $n\sigma$ profile likelihood contours, where $\Delta\chisq  = n^2$, with $n=1$ being the light (blue) shaded region, $n=2$ the dark (blue) shaded region and $n=3$ corresponding to the white area.  The result is compatible with the current LHCb measurement of \g and $\delta_B$, indicated by the point with error bars.  }
	\label{fig:deltaB_vs_gamma}
\end{figure}

The measured observables from the \BToDK decay channels are used to obtain constraints on the underlying physics parameters  $r_B$, $\delta_B$ and \g.  For this purpose, the small effects of $\Dz\Dzb$ mixing and interference in $B^\mp \to D \pi^\mp$ decays are neglected.  Using the measurements and associated fit covariance matrix and systematic uncertainty correlations, and taking external measurements of $\kappa^{K\pi \piz}_{D}$, $F^{\pip \pim \piz}_{\rm +}$ and $F^{\Kp \Km \piz}_{\rm +}$~\cite{Libby:2014rea,Nayak:2014tea} and the branching ratios of the $D$ decay channels~\cite{PDG2014} as additional inputs with their associated uncertainties, a global \chisq minimisation is performed.  A scan of the physics parameters is executed for a range of values and the difference in goodness of fit, $\Delta \chisq$, between the parameter scan values and the global minimum, is evaluated.  Assuming that this \chisq minimisation function is  distributed in a Gaussian manner enables a probability to be assigned  for each set of values of the physics parameters.

Two-dimensional scans are performed for $\gamma$ vs.~$r_B$ and $\gamma$ vs.~$\delta_B$  in the ranges $0.03 < r_B < 0.16$, $0^\circ < \delta_B < 180^\circ$ and $0^\circ < \g < 180^\circ$.  Figs.~\ref{fig:rb_vs_gamma} and \ref{fig:deltaB_vs_gamma} shows the $1\sigma$, $2\sigma$ and $3\sigma$ contours determined from these scans.  It can be seen that the results are compatible with the values obtained from a global analysis of other LHCb measurements sensitive to $\gamma$ at tree level~\cite{LHCb-PAPER-2013-020,*LHCb-CONF-2014-004}, which are also shown. The scans return a best-fit value for the parameter $r_B$ of $0.11 \pm 0.03$.  No useful constraints are obtained for either \g or $\delta_B$.   However, the measurements of the observables are expected to provide improved precision on these parameters when included in a global analysis of all  LHCb $B^\mp \to DK^\mp$ results.

In summary, measurements of \CP asymmetries and related observables  have been performed using \BToDK and \BToDPi decays with an inclusive analysis of the ADS modes $\D \to \Kmp \pipm \piz$ and, for the first time, the quasi-GLW modes $\D \to \pip \pim \piz$ and $\D \to \Kp \Km \piz$.  
The results for the ADS observables are the most precise measurements of these quantities.  
No evidence of \CP violation is obtained with the current experimental precision.  
 First observations have been made of the decays \supBToDPiWithDToKPiPi0 and \BToDPiWithDToKKPi0 , and first evidence is obtained for the mode  \BToDKWithDToKKPi0 .  When analysed in the context of the underlying physics parameters, the results exhibit good consistency with other LHCb measurements.    The measurements will be valuable in improving knowledge of the unitarity triangle angle \g when combined with LHCb results from  \BToDK measurements exploiting other \D decay channels.

%% file: acknowledgements.tex
\section*{Acknowledgements}

\noindent We express our gratitude to our colleagues in the CERN
accelerator departments for the excellent performance of the LHC. We
thank the technical and administrative staff at the LHCb
institutes. We acknowledge support from CERN and from the national
agencies: CAPES, CNPq, FAPERJ and FINEP (Brazil); NSFC (China);
CNRS/IN2P3 (France); BMBF, DFG, HGF and MPG (Germany); INFN (Italy); 
FOM and NWO (The Netherlands); MNiSW and NCN (Poland); MEN/IFA (Romania); 
MinES and FANO (Russia); MinECo (Spain); SNSF and SER (Switzerland); 
NASU (Ukraine); STFC (United Kingdom); NSF (USA).
The Tier1 computing centres are supported by IN2P3 (France), KIT and BMBF 
(Germany), INFN (Italy), NWO and SURF (The Netherlands), PIC (Spain), GridPP 
(United Kingdom).
We are indebted to the communities behind the multiple open 
source software packages on which we depend. We are also thankful for the 
computing resources and the access to software R\&D tools provided by Yandex LLC (Russia).
Individual groups or members have received support from 
EPLANET, Marie Sk\l{}odowska-Curie Actions and ERC (European Union), 
Conseil g\'{e}n\'{e}ral de Haute-Savoie, Labex ENIGMASS and OCEVU, 
R\'{e}gion Auvergne (France), RFBR (Russia), XuntaGal and GENCAT (Spain), Royal Society and Royal
Commission for the Exhibition of 1851 (United Kingdom).

%% file: LHCb_HD_authorlist_2015-02-27.tex
%%%%%%%%%%%%%%%%%%%%%%%%%%%%%%%%%%%%%%%%%%
\centerline{\large\bf LHCb collaboration}
\begin{flushleft}
\small
R.~Aaij$^{38}$, 
B.~Adeva$^{37}$, 
M.~Adinolfi$^{46}$, 
A.~Affolder$^{52}$, 
Z.~Ajaltouni$^{5}$, 
S.~Akar$^{6}$, 
J.~Albrecht$^{9}$, 
F.~Alessio$^{38}$, 
M.~Alexander$^{51}$, 
S.~Ali$^{41}$
% ...

{\footnotesize \it
%$\dagger$ deceased\\[1ex]
% ...
$ ^{5}$Clermont Universit\'{e}, Universit\'{e} Blaise Pascal, CNRS/IN2P3, LPC, Clermont-Ferrand, France\\
$ ^{6}$CPPM, Aix-Marseille Universit\'{e}, CNRS/IN2P3, Marseille, France\\
% ...
$ ^{9}$Fakult\"{a}t Physik, Technische Universit\"{a}t Dortmund, Dortmund, Germany\\
% ...
$ ^{37}$Universidad de Santiago de Compostela, Santiago de Compostela, Spain\\
$ ^{38}$European Organization for Nuclear Research (CERN), Geneva, Switzerland\\
% ...
$ ^{41}$Nikhef National Institute for Subatomic Physics, Amsterdam, The Netherlands\\
% ...
$ ^{46}$H.H. Wills Physics Laboratory, University of Bristol, Bristol, United Kingdom\\
% ...
$ ^{51}$School of Physics and Astronomy, University of Glasgow, Glasgow, United Kingdom\\
$ ^{52}$Oliver Lodge Laboratory, University of Liverpool, Liverpool, United Kingdom\\
% ...
}
\end{flushleft}
%%%%%%%%%%%%%%%%%%%%%%%%%%%%%%%%%%%%%%%%%%

%% file: main.bbl
\ifx\mcitethebibliography\mciteundefinedmacro
\PackageError{LHCb.bst}{mciteplus.sty has not been loaded}
{This bibstyle requires the use of the mciteplus package.}\fi
\providecommand{\href}[2]{#2}
\begin{mcitethebibliography}{10}
\mciteSetBstSublistMode{n}
\mciteSetBstMaxWidthForm{subitem}{\alph{mcitesubitemcount})}
\mciteSetBstSublistLabelBeginEnd{\mcitemaxwidthsubitemform\space}
{\relax}{\relax}

\bibitem{Cabibbo:1963yz}
N.~Cabibbo, \ifthenelse{\boolean{articletitles}}{\emph{{Unitary symmetry and
  leptonic decays}},
  }{}\href{http://dx.doi.org/10.1103/PhysRevLett.10.531}{Phys.\ Rev.\ Lett.\
  \textbf{10} (1963) 531}\relax
\mciteBstWouldAddEndPuncttrue
\mciteSetBstMidEndSepPunct{\mcitedefaultmidpunct}
{\mcitedefaultendpunct}{\mcitedefaultseppunct}\relax
\EndOfBibitem
\bibitem{Kobayashi:1973fv}
M.~Kobayashi and T.~Maskawa, \ifthenelse{\boolean{articletitles}}{\emph{{CP
  violation in the renormalizable theory of weak interaction}},
  }{}\href{http://dx.doi.org/10.1143/PTP.49.652}{Prog.\ Theor.\ Phys.\
  \textbf{49} (1973) 652}\relax
\mciteBstWouldAddEndPuncttrue
\mciteSetBstMidEndSepPunct{\mcitedefaultmidpunct}
{\mcitedefaultendpunct}{\mcitedefaultseppunct}\relax
\EndOfBibitem
\bibitem{Lees:2013zd}
BaBar collaboration, J.~P. Lees {\em et~al.},
  \ifthenelse{\boolean{articletitles}}{\emph{{Observation of direct CP
  violation in the measurement of the Cabibbo-Kobayashi-Maskawa angle $\gamma$
  with $B^\pm\to D^{(*)}K^{(*)\pm}$ decays}},
  }{}\href{http://dx.doi.org/10.1103/PhysRevD.87.052015}{Phys.\ Rev.\
  \textbf{D87} (2013) 052015}, \href{http://arxiv.org/abs/1301.1029}{{\tt
  arXiv:1301.1029}}\relax
\mciteBstWouldAddEndPuncttrue
\mciteSetBstMidEndSepPunct{\mcitedefaultmidpunct}
{\mcitedefaultendpunct}{\mcitedefaultseppunct}\relax
\EndOfBibitem
\bibitem{Trabelsi:2013uj}
K.~Trabelsi, \ifthenelse{\boolean{articletitles}}{\emph{{Study of direct CP in
  charmed B decays and measurement of the CKM angle $\gamma$ at Belle}},
  }{}\href{http://arxiv.org/abs/1301.2033}{{\tt arXiv:1301.2033}}\relax
\mciteBstWouldAddEndPuncttrue
\mciteSetBstMidEndSepPunct{\mcitedefaultmidpunct}
{\mcitedefaultendpunct}{\mcitedefaultseppunct}\relax
\EndOfBibitem
\bibitem{LHCb-PAPER-2013-020}
LHCb collaboration, R.~Aaij {\em et~al.},
  \ifthenelse{\boolean{articletitles}}{\emph{{A measurement of the CKM angle
  $\gamma$ from a combination of $B^\pm \to Dh^\pm$ analyses}},
  }{}\href{http://dx.doi.org/10.1016/j.physletb.2013.08.020}{Phys.\ Lett.\
  \textbf{B726} (2013) 151}, \href{http://arxiv.org/abs/1305.2050}{{\tt
  arXiv:1305.2050}}\relax
\mciteBstWouldAddEndPuncttrue
\mciteSetBstMidEndSepPunct{\mcitedefaultmidpunct}
{\mcitedefaultendpunct}{\mcitedefaultseppunct}\relax
\EndOfBibitem
\bibitem{LHCb-CONF-2014-004}
{LHCb collaboration}, \ifthenelse{\boolean{articletitles}}{\emph{{Improved
  constraints on $\gamma$: CKM2014 update}}, }{} {Sep}, {2014}.
\newblock \href{https://cds.cern.ch/record/1755256}{LHCb-CONF-2014-004}\relax
\mciteBstWouldAddEndPuncttrue
\mciteSetBstMidEndSepPunct{\mcitedefaultmidpunct}
{\mcitedefaultendpunct}{\mcitedefaultseppunct}\relax
\EndOfBibitem
\bibitem{PDG2014}
Particle Data Group, K.~A. Olive {\em et~al.},
  \ifthenelse{\boolean{articletitles}}{\emph{{\href{http://pdg.lbl.gov/}{Review
  of particle physics}}},
  }{}\href{http://dx.doi.org/10.1088/1674-1137/38/9/090001}{Chin.\ Phys.\
  \textbf{C38} (2014) 090001}\relax
\mciteBstWouldAddEndPuncttrue
\mciteSetBstMidEndSepPunct{\mcitedefaultmidpunct}
{\mcitedefaultendpunct}{\mcitedefaultseppunct}\relax
\EndOfBibitem
\bibitem{Atwood:1996ci}
D.~Atwood, I.~Dunietz, and A.~Soni,
  \ifthenelse{\boolean{articletitles}}{\emph{{Enhanced CP violation with $B \to
  K \Dz (\Dzb)$ modes and extraction of the CKM angle \g}},
  }{}\href{http://dx.doi.org/10.1103/PhysRevLett.78.3257}{Phys.\ Rev.\ Lett.\
  \textbf{78} (1997) 3257}, \href{http://arxiv.org/abs/hep-ph/9612433}{{\tt
  arXiv:hep-ph/9612433}}\relax
\mciteBstWouldAddEndPuncttrue
\mciteSetBstMidEndSepPunct{\mcitedefaultmidpunct}
{\mcitedefaultendpunct}{\mcitedefaultseppunct}\relax
\EndOfBibitem
\bibitem{Atwood:2000ck}
D.~Atwood, I.~Dunietz, and A.~Soni,
  \ifthenelse{\boolean{articletitles}}{\emph{{Improved methods for observing CP
  violation in $\Bpm \to K D$ and measuring the CKM phase \g}},
  }{}\href{http://dx.doi.org/10.1103/PhysRevD.63.036005}{Phys.\ Rev.\
  \textbf{D63} (2001) 036005}, \href{http://arxiv.org/abs/hep-ph/0008090}{{\tt
  arXiv:hep-ph/0008090}}\relax
\mciteBstWouldAddEndPuncttrue
\mciteSetBstMidEndSepPunct{\mcitedefaultmidpunct}
{\mcitedefaultendpunct}{\mcitedefaultseppunct}\relax
\EndOfBibitem
\bibitem{Gronau:1990ra}
M.~Gronau and D.~London, \ifthenelse{\boolean{articletitles}}{\emph{{How to
  determine all the angles of the unitarity triangle from $B_{d}^{0} \to D \KS$
  and $B_{s}^{0} \to D\phi$}},
  }{}\href{http://dx.doi.org/10.1016/0370-2693(91)91756-L}{Phys.\ Lett.\
  \textbf{B253} (1991) 483}\relax
\mciteBstWouldAddEndPuncttrue
\mciteSetBstMidEndSepPunct{\mcitedefaultmidpunct}
{\mcitedefaultendpunct}{\mcitedefaultseppunct}\relax
\EndOfBibitem
\bibitem{Gronau:1991dp}
M.~Gronau and D.~Wyler, \ifthenelse{\boolean{articletitles}}{\emph{{On
  determining a weak phase from \CP asymmetries in charged \B decays}},
  }{}\href{http://dx.doi.org/10.1016/0370-2693(91)90034-N}{Phys.\ Lett.\
  \textbf{B265} (1991) 172}\relax
\mciteBstWouldAddEndPuncttrue
\mciteSetBstMidEndSepPunct{\mcitedefaultmidpunct}
{\mcitedefaultendpunct}{\mcitedefaultseppunct}\relax
\EndOfBibitem
\bibitem{LHCb-PAPER-2012-001}
LHCb collaboration, R.~Aaij {\em et~al.},
  \ifthenelse{\boolean{articletitles}}{\emph{{Observation of $CP$ violation in
  $B^\pm \to D K^\pm$ decays}},
  }{}\href{http://dx.doi.org/10.1016/j.physletb.2012.04.060}{Phys.\ Lett.\
  \textbf{B712} (2012) 203}, Erratum
  \href{http://dx.doi.org/10.1016/j.physletb.2012.05.060}{ibid.\
  \textbf{B713} (2012) 351}, \href{http://arxiv.org/abs/1203.3662}{{\tt
  arXiv:1203.3662}}\relax
\mciteBstWouldAddEndPuncttrue
\mciteSetBstMidEndSepPunct{\mcitedefaultmidpunct}
{\mcitedefaultendpunct}{\mcitedefaultseppunct}\relax
\EndOfBibitem
\bibitem{LHCb-PAPER-2012-055}
LHCb collaboration, R.~Aaij {\em et~al.},
  \ifthenelse{\boolean{articletitles}}{\emph{{Observation of the suppressed ADS
  modes $B^\pm \to [\pi^\pm K^\mp\pi^+\pi^-]_D K^\pm$ and $B^\pm \to [\pi^\pm
  K^\mp \pi^+\pi^-]_D \pi^\pm$}},
  }{}\href{http://dx.doi.org/10.1016/j.physletb.2013.05.009}{Phys.\ Lett.\
  \textbf{B723} (2013) 44}, \href{http://arxiv.org/abs/1303.4646}{{\tt
  arXiv:1303.4646}}\relax
\mciteBstWouldAddEndPuncttrue
\mciteSetBstMidEndSepPunct{\mcitedefaultmidpunct}
{\mcitedefaultendpunct}{\mcitedefaultseppunct}\relax
\EndOfBibitem
\bibitem{LHCb-PAPER-2013-068}
LHCb collaboration, R.~Aaij {\em et~al.},
  \ifthenelse{\boolean{articletitles}}{\emph{{A study of CP violation in $B^\pm
  \to D K^\pm$ and $B^\pm \to D \pi^\pm$ decays with $D \to K^0_S K^\pm
  \pi^\mp$ final states}},
  }{}\href{http://dx.doi.org/10.1016/j.physletb.2014.03.051}{Phys.\ Lett.\
  \textbf{B733} (2014) 36}, \href{http://arxiv.org/abs/1402.2982}{{\tt
  arXiv:1402.2982}}\relax
\mciteBstWouldAddEndPuncttrue
\mciteSetBstMidEndSepPunct{\mcitedefaultmidpunct}
{\mcitedefaultendpunct}{\mcitedefaultseppunct}\relax
\EndOfBibitem
\bibitem{LHCb-PAPER-2014-028}
LHCb collaboration, R.~Aaij {\em et~al.},
  \ifthenelse{\boolean{articletitles}}{\emph{{Measurement of CP violation
  parameters in $B^0\to DK^{*0}$ decays}},
  }{}\href{http://dx.doi.org/10.1103/PhysRevD.90.112002}{Phys.\ Rev.\
  \textbf{D90} (2014) 112002}, \href{http://arxiv.org/abs/1407.8136}{{\tt
  arXiv:1407.8136}}\relax
\mciteBstWouldAddEndPuncttrue
\mciteSetBstMidEndSepPunct{\mcitedefaultmidpunct}
{\mcitedefaultendpunct}{\mcitedefaultseppunct}\relax
\EndOfBibitem
\bibitem{LHCb-PAPER-2014-041}
LHCb collaboration, R.~Aaij {\em et~al.},
  \ifthenelse{\boolean{articletitles}}{\emph{{Measurement of the CKM angle
  $\gamma$ using $B^\pm \to D K^\pm$ with $D \to K_S^0\pi^+\pi^-$,
  $K_S^0K^+K^-$ decays}},
  }{}\href{http://dx.doi.org/10.1007/JHEP10(2014)097}{JHEP \textbf{10} (2014)
  097}, \href{http://arxiv.org/abs/1408.2748}{{\tt arXiv:1408.2748}}\relax
\mciteBstWouldAddEndPuncttrue
\mciteSetBstMidEndSepPunct{\mcitedefaultmidpunct}
{\mcitedefaultendpunct}{\mcitedefaultseppunct}\relax
\EndOfBibitem
\bibitem{LHCb-PAPER-2014-017}
LHCb collaboration, R.~Aaij {\em et~al.},
  \ifthenelse{\boolean{articletitles}}{\emph{{Measurement of $CP$ violation and
  constraints on the CKM angle $\gamma$ in $B^\pm \to D K^\pm$ with $D \to
  K_S^0\pi^+\pi^-$ decays}},
  }{}\href{http://dx.doi.org/10.1016/j.nuclphysb.2014.09.015}{Nucl.\ Phys.\
  \textbf{B888} (2014) 169}, \href{http://arxiv.org/abs/1407.6211}{{\tt
  arXiv:1407.6211}}\relax
\mciteBstWouldAddEndPuncttrue
\mciteSetBstMidEndSepPunct{\mcitedefaultmidpunct}
{\mcitedefaultendpunct}{\mcitedefaultseppunct}\relax
\EndOfBibitem
\bibitem{COHERENCE}
D.~Atwood and A.~Soni, \ifthenelse{\boolean{articletitles}}{\emph{{Role of
  charm factory in extracting CKM phase information via $B \to DK$}},
  }{}\href{http://dx.doi.org/10.1103/PhysRevD.68.033003}{Phys.\ Rev.\
  \textbf{D68} (2003) 033003}, \href{http://arxiv.org/abs/hep-ph/0304085}{{\tt
  arXiv:hep-ph/0304085}}\relax
\mciteBstWouldAddEndPuncttrue
\mciteSetBstMidEndSepPunct{\mcitedefaultmidpunct}
{\mcitedefaultendpunct}{\mcitedefaultseppunct}\relax
\EndOfBibitem
\bibitem{Libby:2014rea}
J.~Libby {\em et~al.}, \ifthenelse{\boolean{articletitles}}{\emph{{New
  determination of the $D^{0} \to K^{-} \pi^{+} \pi^{0}$ and $D^{0} \to K^{-}
  \pi^{+} \pi^{+} \pi^{-}$ coherence factors and average strong-phase
  differences}},
  }{}\href{http://dx.doi.org/10.1016/j.physletb.2014.02.032}{Phys.\ Lett.\
  \textbf{B731} (2014) 197}, \href{http://arxiv.org/abs/1401.1904}{{\tt
  arXiv:1401.1904}}\relax
\mciteBstWouldAddEndPuncttrue
\mciteSetBstMidEndSepPunct{\mcitedefaultmidpunct}
{\mcitedefaultendpunct}{\mcitedefaultseppunct}\relax
\EndOfBibitem
\bibitem{NORMLOWREY}
CLEO collaboration, N.~Lowrey {\em et~al.},
  \ifthenelse{\boolean{articletitles}}{\emph{{Determination of the $D^0 \to K^-
  \pi^+ \pi^0$ and $D^0 \to K^-\pi^+\pi^+\pi^-$ coherence factors and average
  strong-phase differences using quantum-correlated measurements}},
  }{}\href{http://dx.doi.org/10.1103/PhysRevD.80.031105}{Phys.\ Rev.\
  \textbf{D80} (2009) 031105}, \href{http://arxiv.org/abs/0903.4853}{{\tt
  arXiv:0903.4853}}\relax
\mciteBstWouldAddEndPuncttrue
\mciteSetBstMidEndSepPunct{\mcitedefaultmidpunct}
{\mcitedefaultendpunct}{\mcitedefaultseppunct}\relax
\EndOfBibitem
\bibitem{Insler:2012pm}
CLEO collaboration, J.~Insler {\em et~al.},
  \ifthenelse{\boolean{articletitles}}{\emph{{Studies of the decays $D^0
  \rightarrow K_S^0K^-\pi^+$ and $D^0 \rightarrow K_S^0K^+\pi^-$}},
  }{}\href{http://dx.doi.org/10.1103/PhysRevD.85.092016}{Phys.\ Rev.\
  \textbf{D85} (2012) 092016}, \href{http://arxiv.org/abs/1203.3804}{{\tt
  arXiv:1203.3804}}\relax
\mciteBstWouldAddEndPuncttrue
\mciteSetBstMidEndSepPunct{\mcitedefaultmidpunct}
{\mcitedefaultendpunct}{\mcitedefaultseppunct}\relax
\EndOfBibitem
\bibitem{Nayak:2014tea}
M.~Nayak {\em et~al.}, \ifthenelse{\boolean{articletitles}}{\emph{{First
  determination of the CP content of $D \to \pi^+ \pi^- \pi^0$ and $D \to
  K^+K^-\pi^0$}},
  }{}\href{http://dx.doi.org/10.1016/j.physletb.2014.11.022}{Phys.\ Lett.\
  \textbf{B740} (2015) 1}, \href{http://arxiv.org/abs/1410.3964}{{\tt
  arXiv:1410.3964}}\relax
\mciteBstWouldAddEndPuncttrue
\mciteSetBstMidEndSepPunct{\mcitedefaultmidpunct}
{\mcitedefaultendpunct}{\mcitedefaultseppunct}\relax
\EndOfBibitem
\bibitem{Lees:2011up}
BaBar collaboration, J.~P. Lees {\em et~al.},
  \ifthenelse{\boolean{articletitles}}{\emph{{Search for $b \to u$ transitions
  in $B^\pm \to [K^\mp \pi^\pm \pi^0]_D K^\pm$ decays}},
  }{}\href{http://dx.doi.org/10.1103/PhysRevD.84.012002}{Phys.\ Rev.\
  \textbf{D84} (2011) 012002}, \href{http://arxiv.org/abs/1104.4472}{{\tt
  arXiv:1104.4472}}\relax
\mciteBstWouldAddEndPuncttrue
\mciteSetBstMidEndSepPunct{\mcitedefaultmidpunct}
{\mcitedefaultendpunct}{\mcitedefaultseppunct}\relax
\EndOfBibitem
\bibitem{Nayak:2013tgg}
Belle collaboration, M.~Nayak {\em et~al.},
  \ifthenelse{\boolean{articletitles}}{\emph{{Evidence for the suppressed decay
  $B^- \to DK^-$, $ D\to K^+\pi^-\pi^0$}},
  }{}\href{http://dx.doi.org/10.1103/PhysRevD.88.091104}{Phys.\ Rev.\
  \textbf{D88} (2013) 091104}, \href{http://arxiv.org/abs/1310.1741}{{\tt
  arXiv:1310.1741}}\relax
\mciteBstWouldAddEndPuncttrue
\mciteSetBstMidEndSepPunct{\mcitedefaultmidpunct}
{\mcitedefaultendpunct}{\mcitedefaultseppunct}\relax
\EndOfBibitem
\bibitem{Aubert:2007ii}
BaBar collaboration, B.~Aubert {\em et~al.},
  \ifthenelse{\boolean{articletitles}}{\emph{{Measurement of CP violation
  parameters with a Dalitz plot analysis of $B^\pm \to D(\pi^+ \pi^{-} \pi^0$ )
  $K^\pm$}}, }{}\href{http://dx.doi.org/10.1103/PhysRevLett.99.251801}{Phys.\
  Rev.\ Lett.\  \textbf{99} (2007) 251801},
  \href{http://arxiv.org/abs/hep-ex/0703037}{{\tt arXiv:hep-ex/0703037}}\relax
\mciteBstWouldAddEndPuncttrue
\mciteSetBstMidEndSepPunct{\mcitedefaultmidpunct}
{\mcitedefaultendpunct}{\mcitedefaultseppunct}\relax
\EndOfBibitem
\bibitem{Rama:2013voa}
M.~Rama, \ifthenelse{\boolean{articletitles}}{\emph{{Effect of $D-\bar{D}$
  mixing in the extraction of $\gamma$ with $B \to D^0 K^-$ and $B^- \to D^0
  \pi^-$ decays}},
  }{}\href{http://dx.doi.org/10.1103/PhysRevD.89.014021}{Phys.\ Rev.\
  \textbf{D89} (2014) 014021}, \href{http://arxiv.org/abs/1307.4384}{{\tt
  arXiv:1307.4384}}\relax
\mciteBstWouldAddEndPuncttrue
\mciteSetBstMidEndSepPunct{\mcitedefaultmidpunct}
{\mcitedefaultendpunct}{\mcitedefaultseppunct}\relax
\EndOfBibitem
\bibitem{Alves:2008zz}
LHCb collaboration, A.~A. Alves~Jr.\ {\em et~al.},
  \ifthenelse{\boolean{articletitles}}{\emph{{The \lhcb detector at the LHC}},
  }{}\href{http://dx.doi.org/10.1088/1748-0221/3/08/S08005}{JINST \textbf{3}
  (2008) S08005}\relax
\mciteBstWouldAddEndPuncttrue
\mciteSetBstMidEndSepPunct{\mcitedefaultmidpunct}
{\mcitedefaultendpunct}{\mcitedefaultseppunct}\relax
\EndOfBibitem
\bibitem{LHCb-DP-2014-002}
LHCb collaboration, R.~Aaij {\em et~al.},
  \ifthenelse{\boolean{articletitles}}{\emph{{LHCb detector performance}},
  }{}\href{http://dx.doi.org/10.1142/S0217751X15300227}{Int.\ J.\ Mod.\ Phys.\
  \textbf{A30} (2015) 1530022}, \href{http://arxiv.org/abs/1412.6352}{{\tt
  arXiv:1412.6352}}\relax
\mciteBstWouldAddEndPuncttrue
\mciteSetBstMidEndSepPunct{\mcitedefaultmidpunct}
{\mcitedefaultendpunct}{\mcitedefaultseppunct}\relax
\EndOfBibitem
\bibitem{LHCb-DP-2012-004}
R.~Aaij {\em et~al.}, \ifthenelse{\boolean{articletitles}}{\emph{{The \lhcb
  trigger and its performance in 2011}},
  }{}\href{http://dx.doi.org/10.1088/1748-0221/8/04/P04022}{JINST \textbf{8}
  (2013) P04022}, \href{http://arxiv.org/abs/1211.3055}{{\tt
  arXiv:1211.3055}}\relax
\mciteBstWouldAddEndPuncttrue
\mciteSetBstMidEndSepPunct{\mcitedefaultmidpunct}
{\mcitedefaultendpunct}{\mcitedefaultseppunct}\relax
\EndOfBibitem
\bibitem{BBDT}
V.~V. Gligorov and M.~Williams,
  \ifthenelse{\boolean{articletitles}}{\emph{{Efficient, reliable and fast
  high-level triggering using a bonsai boosted decision tree}},
  }{}\href{http://dx.doi.org/10.1088/1748-0221/8/02/P02013}{JINST \textbf{8}
  (2013) P02013}, \href{http://arxiv.org/abs/1210.6861}{{\tt
  arXiv:1210.6861}}\relax
\mciteBstWouldAddEndPuncttrue
\mciteSetBstMidEndSepPunct{\mcitedefaultmidpunct}
{\mcitedefaultendpunct}{\mcitedefaultseppunct}\relax
\EndOfBibitem
\bibitem{Sjostrand:2006za}
T.~Sj\"{o}strand, S.~Mrenna, and P.~Skands,
  \ifthenelse{\boolean{articletitles}}{\emph{{PYTHIA 6.4 physics and manual}},
  }{}\href{http://dx.doi.org/10.1088/1126-6708/2006/05/026}{JHEP \textbf{05}
  (2006) 026}, \href{http://arxiv.org/abs/hep-ph/0603175}{{\tt
  arXiv:hep-ph/0603175}}\relax
\mciteBstWouldAddEndPuncttrue
\mciteSetBstMidEndSepPunct{\mcitedefaultmidpunct}
{\mcitedefaultendpunct}{\mcitedefaultseppunct}\relax
\EndOfBibitem
\bibitem{Sjostrand:2007gs}
T.~Sj\"{o}strand, S.~Mrenna, and P.~Skands,
  \ifthenelse{\boolean{articletitles}}{\emph{{A brief introduction to PYTHIA
  8.1}}, }{}\href{http://dx.doi.org/10.1016/j.cpc.2008.01.036}{Comput.\ Phys.\
  Commun.\  \textbf{178} (2008) 852},
  \href{http://arxiv.org/abs/0710.3820}{{\tt arXiv:0710.3820}}\relax
\mciteBstWouldAddEndPuncttrue
\mciteSetBstMidEndSepPunct{\mcitedefaultmidpunct}
{\mcitedefaultendpunct}{\mcitedefaultseppunct}\relax
\EndOfBibitem
\bibitem{LHCb-PROC-2010-056}
I.~Belyaev {\em et~al.}, \ifthenelse{\boolean{articletitles}}{\emph{{Handling
  of the generation of primary events in Gauss, the LHCb simulation
  framework}}, }{}\href{http://dx.doi.org/10.1088/1742-6596/331/3/032047}{{J.\
  Phys.\ Conf.\ Ser.\ } \textbf{331} (2011) 032047}\relax
\mciteBstWouldAddEndPuncttrue
\mciteSetBstMidEndSepPunct{\mcitedefaultmidpunct}
{\mcitedefaultendpunct}{\mcitedefaultseppunct}\relax
\EndOfBibitem
\bibitem{Lange:2001uf}
D.~J. Lange, \ifthenelse{\boolean{articletitles}}{\emph{{The EvtGen particle
  decay simulation package}},
  }{}\href{http://dx.doi.org/10.1016/S0168-9002(01)00089-4}{Nucl.\ Instrum.\
  Meth.\  \textbf{A462} (2001) 152}\relax
\mciteBstWouldAddEndPuncttrue
\mciteSetBstMidEndSepPunct{\mcitedefaultmidpunct}
{\mcitedefaultendpunct}{\mcitedefaultseppunct}\relax
\EndOfBibitem
\bibitem{Golonka:2005pn}
P.~Golonka and Z.~Was, \ifthenelse{\boolean{articletitles}}{\emph{{PHOTOS Monte
  Carlo: A precision tool for QED corrections in $Z$ and $W$ decays}},
  }{}\href{http://dx.doi.org/10.1140/epjc/s2005-02396-4}{Eur.\ Phys.\ J.\
  \textbf{C45} (2006) 97}, \href{http://arxiv.org/abs/hep-ph/0506026}{{\tt
  arXiv:hep-ph/0506026}}\relax
\mciteBstWouldAddEndPuncttrue
\mciteSetBstMidEndSepPunct{\mcitedefaultmidpunct}
{\mcitedefaultendpunct}{\mcitedefaultseppunct}\relax
\EndOfBibitem
\bibitem{Allison:2006ve}
Geant4 collaboration, J.~Allison {\em et~al.},
  \ifthenelse{\boolean{articletitles}}{\emph{{Geant4 developments and
  applications}}, }{}\href{http://dx.doi.org/10.1109/TNS.2006.869826}{IEEE
  Trans.\ Nucl.\ Sci.\  \textbf{53} (2006) 270}\relax
\mciteBstWouldAddEndPuncttrue
\mciteSetBstMidEndSepPunct{\mcitedefaultmidpunct}
{\mcitedefaultendpunct}{\mcitedefaultseppunct}\relax
\EndOfBibitem
\bibitem{Agostinelli:2002hh}
Geant4 collaboration, S.~Agostinelli {\em et~al.},
  \ifthenelse{\boolean{articletitles}}{\emph{{Geant4: A simulation toolkit}},
  }{}\href{http://dx.doi.org/10.1016/S0168-9002(03)01368-8}{Nucl.\ Instrum.\
  Meth.\  \textbf{A506} (2003) 250}\relax
\mciteBstWouldAddEndPuncttrue
\mciteSetBstMidEndSepPunct{\mcitedefaultmidpunct}
{\mcitedefaultendpunct}{\mcitedefaultseppunct}\relax
\EndOfBibitem
\bibitem{LHCb-PROC-2011-006}
M.~Clemencic {\em et~al.}, \ifthenelse{\boolean{articletitles}}{\emph{{The
  \lhcb simulation application, Gauss: Design, evolution and experience}},
  }{}\href{http://dx.doi.org/10.1088/1742-6596/331/3/032023}{{J.\ Phys.\ Conf.\
  Ser.\ } \textbf{331} (2011) 032023}\relax
\mciteBstWouldAddEndPuncttrue
\mciteSetBstMidEndSepPunct{\mcitedefaultmidpunct}
{\mcitedefaultendpunct}{\mcitedefaultseppunct}\relax
\EndOfBibitem
\bibitem{Hulsbergen:2005pu}
W.~D. Hulsbergen, \ifthenelse{\boolean{articletitles}}{\emph{{Decay chain
  fitting with a Kalman filter}},
  }{}\href{http://dx.doi.org/10.1016/j.nima.2005.06.078}{Nucl.\ Instrum.\
  Meth.\  \textbf{A552} (2005) 566},
  \href{http://arxiv.org/abs/physics/0503191}{{\tt
  arXiv:physics/0503191}}\relax
\mciteBstWouldAddEndPuncttrue
\mciteSetBstMidEndSepPunct{\mcitedefaultmidpunct}
{\mcitedefaultendpunct}{\mcitedefaultseppunct}\relax
\EndOfBibitem
\bibitem{Breiman}
L.~Breiman, J.~H. Friedman, R.~A. Olshen, and C.~J. Stone, {\em Classification
  and regression trees}, Wadsworth international group, Belmont, California,
  USA, 1984\relax
\mciteBstWouldAddEndPuncttrue
\mciteSetBstMidEndSepPunct{\mcitedefaultmidpunct}
{\mcitedefaultendpunct}{\mcitedefaultseppunct}\relax
\EndOfBibitem
\bibitem{Friedman2002367}
J.~H. Friedman, \ifthenelse{\boolean{articletitles}}{\emph{Stochastic gradient
  boosting},
  }{}\href{http://dx.doi.org/http://dx.doi.org/10.1016/S0167-9473(01)00065-2}{{Computational
  Statistics \& Data Analysis} \textbf{38} (2002) 367}\relax
\mciteBstWouldAddEndPuncttrue
\mciteSetBstMidEndSepPunct{\mcitedefaultmidpunct}
{\mcitedefaultendpunct}{\mcitedefaultseppunct}\relax
\EndOfBibitem
\bibitem{Pivk:2004ty}
M.~Pivk and F.~R. Le~Diberder,
  \ifthenelse{\boolean{articletitles}}{\emph{{sPlot: A statistical tool to
  unfold data distributions}},
  }{}\href{http://dx.doi.org/10.1016/j.nima.2005.08.106}{Nucl.\ Instrum.\
  Meth.\  \textbf{A555} (2005) 356},
  \href{http://arxiv.org/abs/physics/0402083}{{\tt
  arXiv:physics/0402083}}\relax
\mciteBstWouldAddEndPuncttrue
\mciteSetBstMidEndSepPunct{\mcitedefaultmidpunct}
{\mcitedefaultendpunct}{\mcitedefaultseppunct}\relax
\EndOfBibitem
\bibitem{LHCb-DP-2012-003}
M.~Adinolfi {\em et~al.},
  \ifthenelse{\boolean{articletitles}}{\emph{{Performance of the \lhcb RICH
  detector at the LHC}},
  }{}\href{http://dx.doi.org/10.1140/epjc/s10052-013-2431-9}{Eur.\ Phys.\ J.\
  \textbf{C73} (2013) 2431}, \href{http://arxiv.org/abs/1211.6759}{{\tt
  arXiv:1211.6759}}\relax
\mciteBstWouldAddEndPuncttrue
\mciteSetBstMidEndSepPunct{\mcitedefaultmidpunct}
{\mcitedefaultendpunct}{\mcitedefaultseppunct}\relax
\EndOfBibitem
\bibitem{LHCb-PAPER-2012-052}
LHCb collaboration, R.~Aaij {\em et~al.},
  \ifthenelse{\boolean{articletitles}}{\emph{{Search for $CP$ violation in $D^+
  \to \phi \pi^+$ and $D_s^+ \to K^0_S \pi^+$ decays}},
  }{}\href{http://dx.doi.org/10.1007/JHEP06(2013)112}{JHEP \textbf{06} (2013)
  112}, \href{http://arxiv.org/abs/1303.4906}{{\tt arXiv:1303.4906}}\relax
\mciteBstWouldAddEndPuncttrue
\mciteSetBstMidEndSepPunct{\mcitedefaultmidpunct}
{\mcitedefaultendpunct}{\mcitedefaultseppunct}\relax
\EndOfBibitem
\bibitem{LHCb-PAPER-2014-013}
LHCb collaboration, R.~Aaij {\em et~al.},
  \ifthenelse{\boolean{articletitles}}{\emph{{Measurement of $CP$ asymmetry in
  $D^0 \to K^- K^+$ and $D^0 \to \pi^- \pi^+$ decays}},
  }{}\href{http://dx.doi.org/10.1007/JHEP07(2014)041}{JHEP \textbf{07} (2014)
  041}, \href{http://arxiv.org/abs/1405.2797}{{\tt arXiv:1405.2797}}\relax
\mciteBstWouldAddEndPuncttrue
\mciteSetBstMidEndSepPunct{\mcitedefaultmidpunct}
{\mcitedefaultendpunct}{\mcitedefaultseppunct}\relax
\EndOfBibitem
\end{mcitethebibliography}
